\documentclass{article}
\usepackage{amsfonts}
\usepackage{amsmath}
\usepackage[natbibapa]{apacite}
\usepackage{array}
\usepackage[margin=1in]{geometry}
\usepackage{fancyhdr}
\usepackage[symbol]{footmisc}
\usepackage[pdftex]{graphicx}
\usepackage{mathrsfs}
\usepackage{multirow}
\usepackage{subcaption}
\usepackage{tikz}
\usetikzlibrary{positioning,shapes}

\newcommand{\KLD}[2]{D_{KL}\left( #1 ~ \vert \vert ~ #2 \right)}
\newcommand{\Ent}[1]{H\left( #1 \right)}

\newcommand{\popu}{p_0}
\newcommand{\prior}{p_1}
\newcommand{\ppop}[1]{\popu{}(#1)}
\newcommand{\pprior}[1]{\prior{}(#1)}
\newcommand{\Uprior}{U^1}

\newcommand{\True}[1]{#1_0}
\newcommand{\Sp}[1]{#1_1}

\newcommand{\RV}[1]{{\bf #1}}
\newcommand{\cond}[2]{#1 \vert #2}
\newcommand{\joint}[2]{\left( #1, #2 \right)}

\newcommand{\obsU}{focal divergence}
\newcommand{\ObsU}{Focal divergence}

\DeclareCaptionFormat{apa}
{%
    \textbf{#1#2}#3
}
\DeclareCaptionFormat{apasubfig}
{
    \textbf{#1}#2
}

\newcommand{\notes}[1]{\noindent \emph{Note.} #1}

\begin{document}

\renewcommand*{\thefootnote}{\fnsymbol{footnote}}

\title{Knowing what to know: Implications of the choice of prior distribution on the behavior of adaptive design optimization}

\author{
    Sabina J. Sloman\footnote{
        Corresponding author.
        Please address correspondence to \texttt{sabina.sloman@manchester.ac.uk}, or to Sabina J. Sloman, Department of Computer Science, The University of Manchester, Oxford Rd., Manchester, UK M13 9PL, +44 161 200 8822.
    } ~ \footnote{
        Work undertaken primarily while affiliated with the Department of Social and Decision Sciences, Carnegie Mellon University.
    } \\
    Department of Computer Science, University of Manchester
    \and
    Daniel Cavagnaro \\
    Mihaylo College of Business and Economics, California State University, Fullerton
    \and
    Stephen B. Broomell \\
    Department of Psychological Sciences, Purdue University
}

\maketitle

\renewcommand*{\thefootnote}{\arabic{footnote}}
\setcounter{footnote}{0}

\begin{abstract}
    Adaptive design optimization (ADO) is a state-of-the-art technique for experimental design \citep{cavagnaro_adaptive_2010}. 
    ADO dynamically identifies stimuli that, in expectation, yield the most information about a hypothetical construct of interest (e.g., parameters of a cognitive model). 
    To calculate this expectation, ADO leverages the modeler’s existing knowledge, specified in the form of a prior distribution.
    \emph{Informative} priors align with the distribution of the focal construct in the participant population. 
    This alignment is assumed by ADO's internal assessment of expected information gain. 
    If the prior is instead \emph{misinformative}, i.e., does not align with the participant population, ADO's estimates of expected information gain could be inaccurate.
    In many cases, the true distribution that characterizes the participant population is unknown, and experimenters rely on heuristics in their choice of prior and without an understanding of how this choice affects ADO's behavior.

    Our work introduces a mathematical framework that facilitates investigation of the consequences of the choice of prior distribution on the efficiency of experiments designed using ADO.
    Through theoretical and empirical results, we show that, in the context of \emph{prior misinformation}, measures of expected information gain are distinct from the correctness of the corresponding inference.
    Through a series of simulation experiments, we show that, in the case of parameter estimation, ADO nevertheless outperforms other design methods.
    Conversely, in the case of model selection, misinformative priors can lead inference to favor the wrong model, and rather than mitigating this pitfall, ADO exacerbates it.
\end{abstract}

\section{Introduction}\label{sec:intro}
    Inferences made on the basis of behavioral experiments have the potential to influence both scientific consensus and personalized treatment recommendations.
    However, strong and accurate inferences can require a daunting number of observations, a requirement that can be prohibitive when resources, e.g., participant attention, are scarce.
    Thus, methods that maximize the information provided by each individual observation are extremely valuable.

    \textbf{Adaptive design optimization (ADO)} is a method that leverages observations from individual participants, on-the-fly, to identify the most powerful design in sequence \citep{cavagnaro_adaptive_2010}.\footnote{
      We use the convention that \textbf{terms in bold} refer to definitions, \emph{terms in italics} refer to technical terms that will be defined later, and ``terms in quotations'' refer to vague or ill-defined concepts.
    }
    At its core, ADO evaluates candidate stimuli with a \emph{global utility function} that estimates, for each stimulus, the potential informativeness of possible responses to that stimulus.
    Because of its potential to automatically identify powerful designs, ADO has been used extensively for behavioral, psychometric and psychiatric applications \citep{kwon_adaptive_2022}.
    Such applications are facilitated by the combination of increased access to computational resources and the development of software packages that facilitate its implementation \citep{yang_adopy_2020,sloman_towards_2022}.

    ADO relies on the machinery of Bayesian inference, which requires that the user specify a prior distribution across models and parameter values that will generate their data, i.e., a distribution across possible values of the psychological characteristics underlying the observed stimulus--response relationship.
    When using optimal design methods like ADO, which rely on specified prior distributions in the design of the experiment itself, the choice of prior has dual consequences: Misinformative priors can bias inference and mislead the experimental design process.
    The prior distribution can have a substantial impact on ADO's behavior \citep{myung_tutorial_2013,cavagnaro_functional_2016}.
    Thus, choosing a prior distribution is an issue of enormous practical import, and requires that the experimenter balance multiple considerations, e.g., prior knowledge and analytical tractability \citep{myung_tutorial_2013}.

    The goal of the present work is to unpack the effects of these various considerations on the behavior of ADO.
    We consider a common paradigm in which the goal of the experiment is to measure some latent variable, representing a given psychological characteristic, at the participant level as precisely as possible.
    The assumption is that the behavior exhibited by a given participant can be perfectly captured by a single value of this latent variable, and that these values are drawn from a distribution characterizing the participant population.
    
    In practice, experimenters usually specify a single prior that they use for a large number of experimental participants, their \textbf{specified prior}.
    If the specified prior matches the true distribution of relevant psychological characteristics in the participant population, ADO's criterion for evaluating stimuli can be interpreted as the amount of information the experimenter would receive, on average, across sufficiently many repetitions of the experiment.
    In this case, the design selected by ADO is optimal in the sense that it will lead the experimenter to correct inferences as quickly as possible, on average.
    If the specified prior does not match this population distribution, ADO's global utililty function no longer admits this interpretation, and the designs selected by ADO may no longer lead the experimenter efficiently towards correct inferences.

    Prior literature has devised ways to construct an informed specified prior by incorporating observations from similar past experiments \citep{tulsyan_designing_2012,kim_hierarchical_2014}.
    However, this may be infeasible or impractical in many situations of interest, due to, e.g., resource limitations that restrict the number of total participants one can recruit, or a desire to endow all participants with the same prior knowledge for the sake of ethical considerations or the tractability of pooled analyses.
    In such situations, experimenters are forced to contend with some degree of uncertainty about the true population distribution, and run the risk of deviations between the prior they specify and the population distribution.

    The goal of the present work is to study how deviations between the specified prior and the true population distribution affect the performance of ADO.
    We refer to the presence of such a deviation as \textbf{prior misinformation}.
    In the sections that follow, we introduce a novel conceptual and mathematical framework for investigating the effect of prior misinformation.  
    We leverage this framework to identify both (a) characteristics of specified priors that contribute to robust inference and (b) cases in which the threats of prior misinformation can only be mitigated by acquiring knowledge of the population distribution.
     
    \S \ref{sec:ado} introduces the mechanics of ADO and its application to problems of inference about psychological characteristics, such as trait values and model structure.
    \S \ref{sec:two-lives} presents the main conceptual tension addressed in our paper: Users of ADO implicitly rely on two distinct --- and potentially opposing --- interpretations of the specified prior. 
    \S \ref{sec:efd} gives a mathematical decomposition of the measure of information gain that reveals how prior information affects ADO's efficiency.
    \S \ref{sec:paramest} and \S \ref{sec:modelest} interpret these results in the context of the problems of parameter estimation and model selection, respectively.
    These sections also present results from simulation experiments illustrating the effect of misinformation on the behavior of ADO in practice.
    \S \ref{sec:robust-practices} discusses and suggests practices users of ADO can adopt to enhance robustness to issues we will show can arise in the context of model selection, and \S \ref{sec:discussion} concludes.

\section{Preliminaries}\label{sec:ado}
    \subsection{Notation}
        We use bolded, capital letters to refer to random variables, and lowercase, unbolded letters to refer to their corresponding realizations.
        The probability of a particular realization $x$ of the random variable $\RV{X}$ is $p(x)$, i.e., $\RV{X} ~ : ~ x \rightarrow p(x)$.

    \subsection{Cognitive models}\label{sec:cognitive-models}
        Latent constructs, like those typically of interest in psychological research, are, by definition, unavailable for observation and thus difficult to measure.
        For many applications, experimenters specify cognitive models, which mathematically represent these constructs in such a way that facilitates their measurement.
        The scope of the present work is within-subjects estimation: estimating as precisely as possible the degree to which a given participant exhibits a psychological characteristic.
        We give example applications later in this section.
        First, we make more precise how cognitive models facilitate the measurement of latent psychological constructs.

        We consider probabilistic cognitive models that associate stimuli, e.g., questions that could be asked in an experiment, with probability distributions over possible responses.\footnote{
            In the remainder of this paper, the term ``cognitive model'' can be read as ``probabilistic cognitive model.''
        }
        We denote stimuli $x$ and responses $y$, which are realizations of a random variable $\cond{\RV{Y}}{x}$.
        Models, denoted $m$, are families of functions indexed by a free parameter or parameters, denoted $\theta$.  
        Models encapsulate substantive mechanistic accounts of the relevant psychological, cognitive, or perceptual processes. 
        The parameters encapsulate psychological or behavioral traits that may vary between experimental participants, but which are consistent within a participant.
        Our framework assumes that there is some true model $m^*$ and corresponding parameter value $\theta^*$ that defines the true data-generating distribution for each stimulus $x$, given by $\cond{\mathbf{Y}}{x,\theta^*,m^*}$.
    
        We consider separately the goals of parameter estimation and model selection.
        Parameter estimation is the problem of inferring the value of $\theta^*$, or measuring the degree to which a participant exhibits a particular trait (assuming a given model structure).
        For example, for educational testing, the examiner's goal is to identify the examinee's ability level (assuming a given item-response model).
        Model selection is the problem of inferring the identity of $m^*$ from a set of candidate models $M$, i.e., determining which of several substantively different processes a participant exhibits.  
        For example, a longstanding problem in psychophysics has been to distinguish among various functional forms for describing the relationship between physical dimensions of stimuli and the psychological experience they induce \citep{roberts_measurement_1979}.
        Both of these goals --- parameter estimation and model selection --- can be achieved using Bayesian inference, in which the experimenter places a prior distribution across models and parameter values $\RV{\joint{M}{\Theta}}$ and updates this prior according to observed data.

        By specifying a prior distribution, the experimenter also implicitly specifies a \textbf{prior predictive distribution} $\cond{\RV{Y}}{x}$, for which each possible response to a stimulus has a corresponding marginal probability:

        \begin{align}\label{eq:predictive}
            p(\cond{y}{x}) &= \sum_{m \in M} p(m) \int_\theta p(\cond{y}{x, \theta, m}) ~ p(\cond{\theta}{m}).
        \end{align}

        We can also compute the predictive distribution conditioned on a particular quantity, such as a parameter value or model.

    \subsection{Adaptive design optimization}
        Different sets of stimuli have different degrees of power to identify the generating model and parameter value \citep{myung_optimal_2009,cavagnaro_adaptive_2010,young_rich_2012,broomell_interpreting_2019}.
        To address this, researchers have developed methods for the principled selection of stimuli to maximize the informativeness and efficiency of one's experiment \citep{myung_optimal_2009,broomell_parameter_2014}.
        ADO is one such method \citep{cavagnaro_adaptive_2010}.
        By basing its recommendations on the observations it has seen so far, ADO identifies experimental designs tailored to the response patterns of the current participant.

        Experiments using ADO proceed across a sequence of mini-experiments, which we call trials.  
        Each trial may consist of a single stimulus or a block of stimuli.  
        ADO dynamically incorporates information throughout the experiment by using the posterior distribution from one trial as the prior distribution on the subsequent trial.
        This process is visualized in Figure \ref{fig:ado}.

        \newcommand{\arrowlength}{.6in}
        \newcommand{\arrowbend}{20}

	\begin{figure}
            \caption{}
            \label{fig:ado}
		\begin{center}
			\begin{tikzpicture}
				\node (specify) [fill={rgb:black,1;white,4},rounded corners,text centered] {Specify $\RV{\joint{M}{\Theta}}$};
				\node (select) [fill={rgb:black,1;white,2},rounded corners,above right=\arrowlength and \arrowlength of specify,text centered,text width=1.75 in] {Select $x^* = \mathrm{argmax}_x U$};
				\node (collect) [fill={rgb:black,1;white,2},rounded corners,below right=\arrowlength and \arrowlength of select,text centered] {Collect data $\cond{y}{x^*}$};
				\node (update) [fill={rgb:black,1;white,2},rounded corners,below right=\arrowlength and \arrowlength of specify,text centered,text width=1.75 in] {Update to $\cond{\RV{\joint{M}{\Theta}}}{\{y, x^*\}}$};

				\path[->,bend left=\arrowbend,ultra thick] (specify) edge (select);
				\path[->,bend left=\arrowbend,ultra thick] (select) edge (collect);
				\path[->,bend left=\arrowbend,ultra thick] (collect) edge (update);
				\path[->,bend left=\arrowbend,ultra thick] (update) edge (select);
			\end{tikzpicture}
		\end{center}
		\notes{
                Flow chart of ADO experiment.
                The experimenter begins the experiment at the lightest grey node, by specifying a prior distribution over models and parameter values.
                On each trial, they select the stimulus that maximizes the global utility, observe responses to that stimulus, update the distribution over models and parameter values according to Bayes' rule, and then use the obtained posterior as the prior on the next trial.
            }
	\end{figure}

        To identify the stimulus with the greatest information gain, users of ADO specify a \textbf{local utility function} $u(x, y, \theta, m)$ which is a function of the candidate stimulus $x$, response $y$, and a possible model and parameter value $\{ m, \theta \}$ (together, a possible \textbf{state of the world}).
        The local utility function measures how much is learned from response $y$ on stimulus $x$ about the state of the world $\{ m, \theta \}$.
        It can take a variety of forms, depending on the particular goals of the experimenter.
        The true state of the world and outcome of the experiment are unknown to the experimenter \emph{a priori} --- otherwise, there would be no need to run the experiment.
        Therefore, rather than maximizing $u$, ADO selects the stimulus that maximizes the expectation of $u$ across possible models, parameter values and experimental outcomes according to the specified prior distribution.
        This yields the \textbf{global utility function}:

        \begin{align}\label{eq:U}
            U(x) &= \sum_{m \in M} p(m) \int_\theta \int_y u(x, y, \theta, m) ~ p(y \vert x, \theta, m) ~ p(\theta \vert m).
        \end{align}
        
        For our applications, we consider a specification of $u$ such that Equation \ref{eq:U} measures the amount of information the candidate stimulus $x$ is expected to yield about some inferential quantity of interest.
        The amount of information one variable provides about another has been made mathematically precise in the field of information theory by the concept of mutual information \citep{cover_elements_1991}.
        Motivated by these information-theoretic principles, \textbf{global mutual information utility} is the mutual information ($I$) between a focal quantity of interest, which we refer to as the \textbf{focus} and denote $\phi$, and responses to a stimulus \citep{bernardo_expected_1979}.  
        Then, the global mutual information utility of a stimulus is:\footnote{
            If $\RV{\Phi}$ is a discrete random variable, as is the case in the problem of model selection (\S \ref{sec:mi-modelest}), the integrals in Equation \ref{eq:Umi} are replaced by the analogous sums.
        }

        \begin{align}\label{eq:Umi}
            U(x) &= \int_\phi \int_y \log{\left( \frac{p(\phi \vert y, x)}{p(\phi)} \right)} ~ p(y \vert x, \phi) ~ p(\phi) \nonumber \\
            &= I(\RV{\Phi} ; \mathbf{Y} \vert x).
        \end{align}

        In order for the global utility function to have the form in Equation \ref{eq:Umi}, the local utility function must take the form:

        \begin{align}
            u(x, y, \theta, m) &= \log{\left( \frac{p(\phi \vert y, x)}{p(\phi)} \right)}
        \end{align}
        which can be thought of as a measure of the information gained about the true value of $\phi$ from $\cond{y}{x}$.

        \S \ref{sec:mi-paramest} and \S \ref{sec:mi-modelest} show how this specification is adapted to two of the most frequent applications of ADO: the problems of parameter estimation and of model selection.
        In the former case, the parameters $\theta$ are the focus, and in the latter case, the model $m$ is the focus.

        Notice that Equation \ref{eq:Umi} can be rewritten in terms of Kullback-Leibler divergence, an information-theoretic measure that captures the information gained in moving from one distribution to another.  
        Specifically:

        \begin{align}\label{eq:Udiverg}
            U(x) &= \int_y \underbrace{\KLD{\cond{\RV{\Phi}}{y,x}}{\RV{\Phi}}}_{\text{\ObsU{}}} ~ p(y \vert x)
        \end{align}
        where $\KLD{\cond{\RV{\Phi}}{y,x}}{\RV{\Phi}}$, or what we will refer to as the \textbf{focal divergence}, is the Kullback-Leibler divergence from distribution $\cond{\RV{\Phi}}{x,y}$ to distribution $\RV{\Phi}$.
        In other words, global mutual information utility captures, in an information-theoretic sense, how much an observed response to a particular stimulus $x$ is expected to move the prior distribution assigned to the focus.

    \subsection{Parameter estimation}\label{sec:mi-paramest}
        Parameter estimation refers to the problem of maximizing the precision of one's estimate of the parameters $\theta$ given a particular model $m$.
        Applications of ADO to parameter estimation are useful if the experimenter is interested in capturing individual variation, for the purpose of, e.g., generating personalized treatment recommendations on the basis of a behavioral assessment.
        In the educational testing setting mentioned above, the examiner's goal is to identify each examinee's ability level in order to make recommendations of areas of strength or potential improvement \citep{owen_bayesian_1969}.
        In a medical application, \citet{hou_evaluating_2016} used ADO to estimate participants' degree of visual contrast sensitivity, a characteristic that can be used for diagnosis of eye disease and treatment recommendations.

        In the context of ADO for parameter estimation, $m$ is assumed known, and the focus of the utility function is the parameter $\theta$.
        The global utility function is:

        \begin{align}
            U(x) &= \int_\theta \int_y \log{\left( \frac{p(\theta \vert y, x)}{p(\theta)} \right)} ~ p(y \vert \theta) ~ p(\theta).
        \end{align}

        \paragraph{Focal predictive distributions}
            As mentioned in \S \ref{sec:cognitive-models}, we can compute the predictive distribution conditioned on any particular state of the world, $\cond{\RV{Y}}{x, \theta, m}$ (Equation \ref{eq:predictive}).
            In the context of parameter estimation, the value of $m$ is known by assumption, so we can equivalently compute the predictive distribution conditioned on any particular value of $\theta$, $\cond{\RV{Y}}{x, \theta}$.
            In this case, the set of predictive distributions characterized by possible parameter values are also the set of \textbf{focal predictive distributions}, or the predictive distributions associated with possible values of the focus.

            We highlight two properties of the focal predictive distributions in the context of parameter estimation.
            First, since the true data-generating distribution is $\cond{\RV{Y}}{x, \theta}$ for some value of $\theta$, the set of focal predictive distributions is in effect a set of possible data-generating distributions.
            The parameter estimation problem then (asymptotically) amounts to identifying which value of the focus has a corresponding predictive distribution that most resembles the distribution of observed data.
            
            Second, because of this, the predictive distribution corresponding to a particular value of the focus does not depend on additional information like the current trial number or history of observations: While a particular value of $\theta$ may become arbitrarily more or less likely, it will always elicit the same likelihood on a given stimulus--response pair.

    \subsection{Model selection}\label{sec:mi-modelest}
        Model selection refers to the problem of maximizing the precision of one's estimate of the model $m$, assuming both $m$ and $\theta$ are unknown.
        The problem of model selection can be thought of as identifying the core psychological process governing a participant's response distribution.

        In the context of model selection, the focus of the utility function is the model $m$, which yields the global utility function:

        \begin{align}\label{eq:Umodel}
            U(x) &= \sum_{m \in M} p(m) \int_y \log{\left( \frac{p(m \vert y, x)}{p(m)} \right)} ~ p(y \vert x, m) \nonumber \\
            &= \sum_{m \in M} p(m) \int_\theta \int_y \log{\left( \frac{p(m \vert y, x)}{p(m)} \right)} ~ p(y \vert x, \theta, m) ~ p(\theta \vert m).
        \end{align}

        \paragraph{Focal predictive distributions}
            In the case of model selection, the focal predictive distributions are the predictive distributions associated with possible values of the model $m$, which can be calculated as:

            \begin{align}\label{eq:predictive-model}
                p(\cond{y}{x, m}) &= \int_\theta p(\cond{y}{x, \theta, m}) ~ p(\cond{\theta}{m}).
            \end{align}

            Experimenters faced with the model selection problem have two sources of uncertainty to contend with (the value of $m$ and the value of $\theta$), yet measure utility with respect to reduction in only one source of uncertainty.
            This is reflected in properties of Equation \ref{eq:predictive-model}: Unlike in the case of parameter estimation, here, the focus is not the only conditioning variable needed to completely specify a possible response distribution $\cond{\RV{Y}}{x,\theta,m}$; full specification of the response distribution also requires knowledge of $\theta$.\footnote{
                For this reason, the problem of model selection is a special case of an embedded model problem \citep{foster_variational_2021}, or inference in the presence of nuisance parameters \citep{paninski_asymptotic_2005}.
            }
            In addition, unlike in the case of parameter estimation, the focal predictive distributions are a moving target: Because of their dependence on the parameter distributions, they shift as the parameter distributions are updated on the basis of observed data.
            These characteristics will become important in our discussion in \S \ref{sec:modelest} of the impact of prior misinformation in the context of model selection.

\section{The prior's two lives}\label{sec:two-lives}
    In ADO, the specified prior plays two roles: It both facilitates estimation of the focus from data via Bayesian updating, and informs the design of the experiment that generates these data.
    These two roles, or ``lives,'' of the prior map on to two traditions in Bayesian statistics: Bayesian inference and Bayesian decision theory.
    While the effect of the prior on the behavior of Bayesian inference has been well-studied, specified priors that enjoy good theoretical guarantees in the context of Bayesian inference may not seem so appealing when evaluated on the quality of a corresponding sequential decision-making policy.
    This section unpacks the reasons for this.
    The goal of the present work is, in a sense, parallel to that of literature understanding the effect of priors on Bayesian inference: Our goal is to understand the effect of the choice of prior distribution on the quality of the corresponding sequential decision-making policy, and give guidance for users of ADO constrained to identify a single prior that lives both lives.

    Sequential Bayesian inference is a core component of ADO: On each trial, the prior distribution is constructed as the posterior from the previous trial.
    In its first role, the prior can be seen as a launching pad for learning that will occur throughout the experiment.
    The prior is understood as an incomplete and ill-informed characterization of the distribution over possible states of the world, and is usually constructed on the basis of a variety of epistemic and pragmatic considerations.
    Considerations pertaining to --- and guidance for constructing --- the prior in the context of sequential Bayesian inference is the topic of a substantial body of existing literature (e.g., \citet{lopes_confronting_2011,gelman_prior_2017}).
    \emph{Uninformative} priors are often selected because of their pragmatic appeal in this role.

    In its second role, the prior is used when calculating the global utility (Equation \ref{eq:U}) and thus informs the experimental design policy about the relative likelihoods of various outcomes.
    Bayesian decision theory refers to a prescriptive decision-making policy in which the costs and benefits of taking an action in different states of the world are averaged according to the probabilities of those states of the world \citep{degroot_optimal_2005,berger_statistical_2013}.
    ADO's policy of selecting the stimulus that maximizes the global utility is a special case of a Bayesian decision theoretic method.
    If the decision-making policy relies on a prior that mischaracterizes the relative likelihoods of candidate states of the world, the prescribed action is no longer defensible as the action with the highest expected benefit.
    Bayesian decision theoretic applications thus require a prior that is as informed as possible with available knowledge about the distribution of states of the world.
    Priors that ignore or mislead about the available knowledge can not be easily justified from a decision-theoretic perspective, as they may bias the design selection toward stimuli that would not actually be the most informative across multiple experiments.

    We assume that the relevant prior knowledge is the true distribution of relevant psychological characteristics in the participant population.
    Therefore, we will refer to the best decision-theoretic prior as the \textbf{population prior}.
    We do this for conceptual tractability; however, the analyses that follow require only that there is some defensible decision-theoretic prior. 
    Our results apply regardless of the basis on which that prior is constructed.
    In many cases, information in addition to or instead of a population distribution should inform the decision-theoretic prior.
    For example, in all but the first trial of an adaptive experiment, the decision-theoretic prior must condition on the observations seen in previous trials.
    In these cases, the decision-theoretic prior can be formed from the population distribution conditioned on the history of observations (our analyses incorporate this consideration, in a way that is stated more formally in \S \ref{sec:extended-notation}).
    More generally, our framework extends to any case in which other information, e.g., knowledge about relevant demographic characteristics or a participant's past behavior, is available, as our results can be readily generalized by considering the ``population'' as all participants with the same demographic or behavioral characteristics.
    
    If the specified prior --- the prior used in the context of the experiment --- matches the population prior, the global utility (Equation \ref{eq:Umi}) is also the \emph{expected \obsU{}} --- the degree of \obsU{} one should expect if one were to run the experiment on a sufficiently large participant sample.
    On the other hand, if the specified prior is not well-calibrated, the global utility values could be misleading about the expected \obsU{}.
    \S \ref{sec:motivation} gives an example of this in the context of an item-response model, a common paradigm used for educational testing.
    Experiments identified by ADO may not have the power to precisely identify the true model or its parameters, leading to a situation where a characteristic indicative of a disease or needed intervention is not identified efficiently, or possibly at all.

    \subsection{Types of priors}\label{sec:types-of-priors}
        Priors are typically categorized as ``informative'' or ``uninformative.''  
        With an informative prior, a Bayesian analysis may reach a different conclusion than a conventional one because the prior injects information that is not in the data.  
        For a single experiment aimed at identifying the model and parameter of an individual, the ideal informative prior would be a degenerate one that gives probability 1 to the true model and parameter.  
        Such a prior is not feasible for the paradigm we consider here, where the same prior must be used for each participant drawn from a heterogeneous population.  
        For this case, the best one could do would be to use a population prior.  
        The logic of ADO implicitly assumes that the specified prior is the population prior.  
        Therefore, we characterize the prior that coincides with the population prior as \textbf{informative}, and any prior that deviates from that population prior as \textbf{misinformative}.

        Under our definition, priors that are usually referred to as ``uninformative'' are typically misinformative when considered in the context of decision-theoretic applications.
        ``Uninformative'' priors are not supposed to inject information, but in the paradigm we consider here, they entail explicit assumptions about the population of participants in the study.
        We will here use \textbf{uninformative} in the context of parameter estimation to refer to a special class of misinformative priors that are agnostic about either the parameter value or the predictive distribution.
        Priors that are agnostic about the parameter value --- are \textbf{uninformative in parameter space} --- are disperse across the support of the parameter distribution.
        Priors that are agnostic about the data distribution --- are \textbf{uninformative in data space} --- have high density in regions of the parameter space that correspond to a wide variety of data distributions.
        These two properties do not necessarily, or even usually, coincide.

\section{Expected \obsU{}}\label{sec:efd}
    The primary innovation of our analysis is to decouple the two lives of the prior, and provide a framework within which one can reason separately about the process of sequential Bayesian inference and the distribution of observations upon which this inference is performed.\footnote{
        See \citet{simchowitz_bayesian_2021} for a related analysis in the context of Bayesian decision-making algorithms more generally.
    }
    In this section, we more precisely define, motivate, and mathematically unpack the expected focal divergence, a concept that is central to the remainder of our analyses.

    \subsection{Extended notation}\label{sec:extended-notation}
        In the remainder of our paper, it will be important to distinguish whether a random variable is distributed according to the population or specified distribution of the corresponding quantity.
        We will do this by subscripting variables that correspond to the population distribution with a 0, e.g., the population distribution of models and parameters becomes $\True{\RV{\joint{M}{\Theta}}}$, and the corresponding marginal distribution of observations becomes $\cond{\True{\RV{Y}}}{x}$.
        Analogously, we will subscript variables that correspond to the specified distribution with a 1, e.g., the specified distribution of models and parameters becomes $\Sp{\RV{\joint{M}{\Theta}}}$, and the corresponding marginal distribution of observations, i.e., the distribution of observations implied by the specified prior, becomes $\cond{\Sp{\RV{Y}}}{x}$.
        We will also use $p_0$ and $p_1$ analogously to refer to the probabilities of the implied random variables taking particular values under the true and specified distribution, respectively.
        
        The notation for quantities used repeatedly is summarized in Table \ref{tab:notation}.
        While Table \ref{tab:notation}, and our discussion more generally, refers to prior distributions, i.e., the distributions of random variables before conditioning on observations, all distributions should be interpreted to implicitly condition on the number of observations implied by context.
        For example, we write $\Sp{\RV{\joint{M}{\Theta}}}$ to refer generally to the specified prior, regardless of how many experimental trials have elapsed.
        When considering the degree of prior misinformation on the second trial of an experiment, i.e., after an observation $(x, y)$, this can be read as $\cond{\Sp{\RV{\joint{M}{\Theta}}}}{\{x, y\}}$ (recalling that the posterior from the first trial is the prior on the second trial).
        In the same way, the population posterior distribution is $\cond{\True{\RV{\joint{M}{\Theta}}}}{\{ y,x \}}$, which can be interpreted as the appropriate decision-theoretic prior for the next trial given the history of observations.

        \begin{table}[]
            \caption{}
            \label{tab:notation}
            \emph{Extended Notational System}

            \vspace{4mm}
            
            \hspace{-.5in}
            \begin{tabular}{m{0.3\textwidth}>{\centering}m{0.12\textwidth}>{\centering}m{0.12\textwidth}>{\centering}m{0.4\textwidth}>{\centering\arraybackslash}m{0.1\textwidth}}
                \hline
                \hspace{.7in} \textbf{Terminology} & \textbf{Variable} & \textbf{Realization} & \textbf{Evaluation} & \textbf{Known?} \\
                \hline
                Candidate stimulus & & $x$ & Specified by experimenter & $\checkmark$ \\
                \hline
                Population prior & $\True{\RV{\joint{M}{\Theta}}}$ & \multirow{2}{*}{$\joint{m}{\theta}$} & Property of the system under study & $\times$ \\
                Specified prior & $\Sp{\RV{\joint{M}{\Theta}}}$ & & Specified by experimenter & $\checkmark$ \\
                \hline
                Population distribution of focus & $\True{\RV{\Phi}}$ & \multirow{2}{*}{$\phi$} & Subspace of $\True{\RV{\joint{M}{\Theta}}}$ & $\times$ \\
                Specified distribution of focus & $\Sp{\RV{\Phi}}$ & & Subspace of $\Sp{\RV{\joint{M}{\Theta}}}$ & $\checkmark$ \\
                \hline
                Response distribution & $\cond{\True{\RV{Y}}}{x}$ & \multirow{3}{*}{$y$} & $y \rightarrow \sum_{m \in M} \ppop{m} \int_\theta p(\cond{y}{x, \theta, m}) ~ \ppop{\cond{\theta}{m}}$ & $\times$ \\
                Prior predictive distribution & $\cond{\Sp{\RV{Y}}}{x}$ & & $y \rightarrow \sum_{m \in M} \pprior{m} \int_\theta p(\cond{y}{x, \theta, m}) ~ \pprior{\cond{\theta}{m}}$ & $\checkmark$ \\
                Focal predictive distribution & $\cond{\Sp{\RV{Y}}}{x, \phi}$ & & $y \rightarrow \pprior{\cond{y}{x, \phi}}$ & $\checkmark$ \\
                \hline
                Global utility & & $U(x)$ & $\int_y \int_\phi \log{\left( \frac{\pprior{\cond{\phi}{y, x}}}{\pprior{\phi}} \right)} ~ \pprior{\cond{\phi}{y, x}} ~ \pprior{\cond{y}{x}}$ & $\checkmark$ \\
                Expected focal divergence & & $\Uprior{}(x)$ & $\int_y \int_\phi \log{\left( \frac{\pprior{\cond{\phi}{y, x}}}{\pprior{\phi}} \right)} ~ \pprior{\cond{\phi}{y, x}} ~ \ppop{\cond{y}{x}}$ & $\times$ \\
                \hline
            \end{tabular}

            \vspace{8mm}
            
            \notes{Columns show, respectively, the terminology used for quantities repeatedly referred to, and the corresponding random variable notation, notation used for realizations of the corresponding random variable, how the corresponding distribution is evaluated, and whether the corresponding distribution is available to the experimenter.
            }
        \end{table}

    \subsection{Definition of expected \obsU{}}\label{sec:focal-divergence}
        Equation \ref{eq:Udiverg} showed that global mutual information utility can be rewritten as an expectation of the \obsU{} across the specified predictive distribution.
        In the context of the prior's two lives, the \obsU{} can be thought of as the degree to which the prior fulfills its role of efficient Bayesian inference.
        Taking the expectation of the \obsU{} across the specified predictive distribution then invokes the prior's decision-theoretic role: One uses the predictive distribution implied by the specified prior to calculate the relative likelihood of prospective observations.
        
        In the case where the specified prior deviates from the population prior, i.e., the specified prior is misinformative, the global mutual information utility is not equivalent to the \obsU{} an experimenter would achieve from a stimulus if they presented it to many members of the participant population.
        We refer to this latter quantity --- the expectation of the \obsU{} taken across the response distribution --- as the \textbf{expected \obsU{}}.
        The expected \obsU{} associated with a stimulus $x$, denoted $\Uprior{}(x)$, is:

        \begin{align}\label{eq:Up}
            \Uprior{}(x) &= \int_y \int_\phi \log{\left( \frac{\pprior{\phi \vert y, x}}{\pprior{\phi}} \right)} ~ \pprior{\phi \vert y, x} ~ \ppop{y \vert x} \nonumber \\
            &= \int_y \KLD{\cond{\Sp{\RV{\Phi}}}{\{ y,x \}}}{\Sp{\RV{\Phi}}} ~ \ppop{y \vert x},
        \end{align}
        i.e., is the expected Kullback-Leibler divergence between posterior and prior under the response distribution $\cond{\True{\RV{Y}}}{x}$, or how much observations distributed according to the population distribution are expected to move the prior distribution.

    \subsection{Motivating example}\label{sec:motivation}
        To illustrate our claim that misinformative priors can impact the effectiveness of ADO, we demonstrate how the population distribution can affect the expected \obsU{} of a stimulus in the context of a simple item-response model.
        We consider an item-response model that uses a one-dimensional ``proficiency'' trait $\theta$ to predict the probability of a correct response to a multiple-alternative question with a given ``item difficulty,'' $x$.
        For a fixed value of $x$, higher values of $\theta$, i.e., greater proficiency, yields a higher probability of a correct response.
        For a fixed value of $\theta$, higher values of $x$, i.e., more difficult items, yield lower probabilities of a correct response, with the lowest possible probability being some value greater than zero that is consistent with random guessing.
        The goal of an experiment is to estimate the proficiency of each participant from their responses to items of various difficulty levels.

        In prior work, \citet{weiss_biased_1983} found that priors that differed from the population distribution induced biases in inferences drawn from experiments designed using a version of ADO.\footnote{
            Unlike us, \cite{weiss_biased_1983} did not provide analytical results, did not extend their analysis beyond item-response models, and did not examine the effect of general properties of prior distributions (e.g., dispersion).
        }
        As our running example, we adopt the item-response model used in their simulation study:\footnote{
            We set the item discrimination parameter to the middle of the range investigated by \citet{weiss_biased_1983}, resulting in the constant 2.72 present in Equation \ref{eq:irt}.
        }

        \begin{align}\label{eq:irt}
            p(y = 1 \vert x, \theta) &= .2 + \frac{.8}{1 + e^{-2.72(\theta - x)}}.
        \end{align}

\begin{figure}
            \caption{}
            \label{fig:irt-motivation}
            \begin{subfigure}{.48\linewidth}
                \caption{}
                \label{fig:irt-motivation-pred}
                \includegraphics[width=\linewidth]{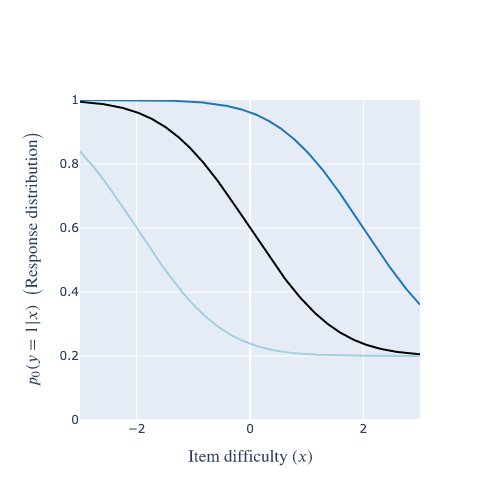}
            \end{subfigure}\hfill\begin{subfigure}{.48\linewidth}
                \caption{}
                \label{fig:irt-motivation-u1}
                \includegraphics[width=\linewidth]{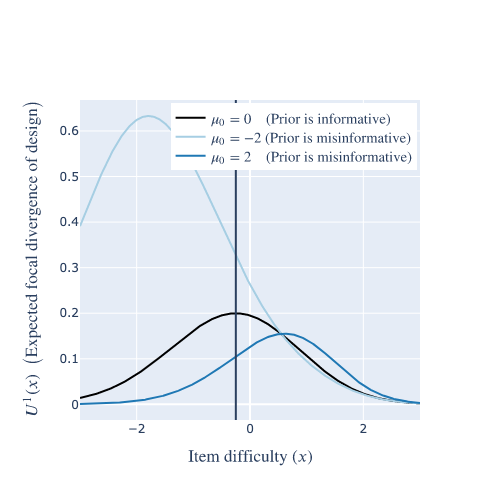}
            \end{subfigure}
            \notes{The effect of prior misinformation: Motivating example from item response theory.
                Effect of the population distribution on \textbf{(a)} response distribution $\ppop{y = 1 \vert x}$, and \textbf{(b)} the expected \obsU{} of a stimulus $\Uprior{}(x)$.
                Colors denote different true distributions $\True{\RV{\Theta}}$.
                In all cases, the specified prior is $\Sp{\RV{\Theta}} \sim \mathscr{N}(0, 1)$ (i.e., is a standard normal distribution).
                The vertical line indicates the stimulus, i.e., value of $x$, that would be selected by ADO under the specified prior.
            }
        \end{figure}

        The black curve in Figure \ref{fig:irt-motivation-pred} shows, for each item difficulty $x$ between -3 and 3, the predictive distribution associated with a prior $\Sp{\RV{\Theta}} \sim \mathscr{N}(0,1)$ (i.e., distributed according to a standard normal distribution).
        In the case this prior is informative, i.e., the population distribution is also $\True{\RV{\Theta}} \sim \mathscr{N}(0,1)$, this curve also shows the empirical distribution of responses one should expect.
        The black curve in Figure \ref{fig:irt-motivation-u1} shows the global utility corresponding to each candidate design under this prior.
        In the case this prior is informative, this curve also shows the expected \obsU{} corresponding to each candidate design.

        The blue curves show the distribution of observations and expected \obsU{} values under two other possible population priors.
        The light blue curves correspond to the population prior $\True{\RV{\Theta}} \sim \mathscr{N}(-2, 1)$, and the dark blue curves correspond to the population prior $\True{\RV{\Theta}} \sim \mathscr{N}(2, 1)$.
        With reference to Figure \ref{fig:irt-motivation-u1}, if the true population distribution is $\True{\RV{\Theta}} \sim \mathscr{N}(2, 1)$, the stimulus selected by ADO will yield much less \obsU{} than ADO anticipates, on average.
        By contrast, if the population distribution is $\True{\RV{\Theta}} \sim \mathscr{N}(-2, 1)$, the stimulus selected by ADO will yield much more \obsU{} than ADO anticipates, on average.

        What accounts for this difference?
        Are there systematic properties of prior distributions that determine which will yield a greater or less expected \obsU{}?
        The following section unpacks these questions.

    \subsection{Decomposition of the expected \obsU{}}\label{sec:uprior}
        The expected \obsU{} $\Uprior{}(x)$ decomposes into three terms, which provide insight into how prior misinformation may affect ADO's efficiency:
        \begin{align}\label{eq:Ulearning}
            \Uprior{}(x) = &~~~~ \Ent{\cond{\True{\RV{Y}}}{x}} &~ \biggr\} ~ &\text{Response variability} \nonumber \\
            &+ \KLD{\cond{\True{\RV{Y}}}{x}}{\cond{\Sp{\RV{Y}}}{x}} &~ \biggr\} ~ &\text{Surprisal} \nonumber \\
            &+ \int_y \int_\phi \log{\left( \pprior{y \vert x, \phi} \right)} ~ \pprior{\phi \vert y, x} ~ \ppop{y \vert x}. &~ \biggr\} ~  &\text{Hindsight}
        \end{align}
        Derivation is deferred to Appendix \ref{ap:U1}.

        \textbf{Response variability} is the entropy in responses to a given stimulus.  
        Entropy is an information-theoretic measure of the uncertainty related to the possible outcomes of a random variable.  
        If a random variable has only one possible outcome then it has no entropy, while a distribution with high entropy is very disperse across its support.  
        This captures the intuitive notion that questions are less informative when the the experimenter already knows what the response will be.
        Response variability stems from a) uncertainty about the value of the focus, and b) uncertainty about the responses given a particular value of the focus.
        The source of the stochasticity will determine how this term affects inference, which we discuss more in \S \ref{sec:param-experiments-irt}.
        Another important characteristic of response variability is that it is a function only of the response distribution, and so should not affect one's choice of prior.
    
        \textbf{Surprisal} is the Kullback-Leibler divergence between the specified prior predictive distribution and the response distribution.  
        Higher surprisal contributes to higher expected \obsU{} since the specified prior is forced to update in light of observed inconsistencies.
        Considered differently, high surprisal indicates that there is a lot to learn --- i.e., the specified prior is in a sense more misinformed.
        Thus, despite its contribution to the expected \obsU{}, one would generally prefer a specified prior that induces low surprisal.
    
        \textbf{Hindsight} is the expected posterior log likelihood of responses under the specified prior.
        Posterior likelihood is a function of both prior likelihood and the specified prior's ability to ``respond'' to observations.
        We discuss this property of ``responsiveness'' more formally in \S \ref{sec:paramest}.
        Surprisal and hindsight will tend to be inversely related through the prior likelihood.
        Our discussion of considerations in the specification of one's prior, particularly in \S \ref{sec:paramest}, will focus on the effect of different specified priors on hindsight.

    \subsection{Revisiting motivating example}\label{sec:motivation-unpacked}
\begin{figure}
            \caption{}
            \label{fig:irt-learningu_priorfixed}
            \includegraphics[width=\linewidth]{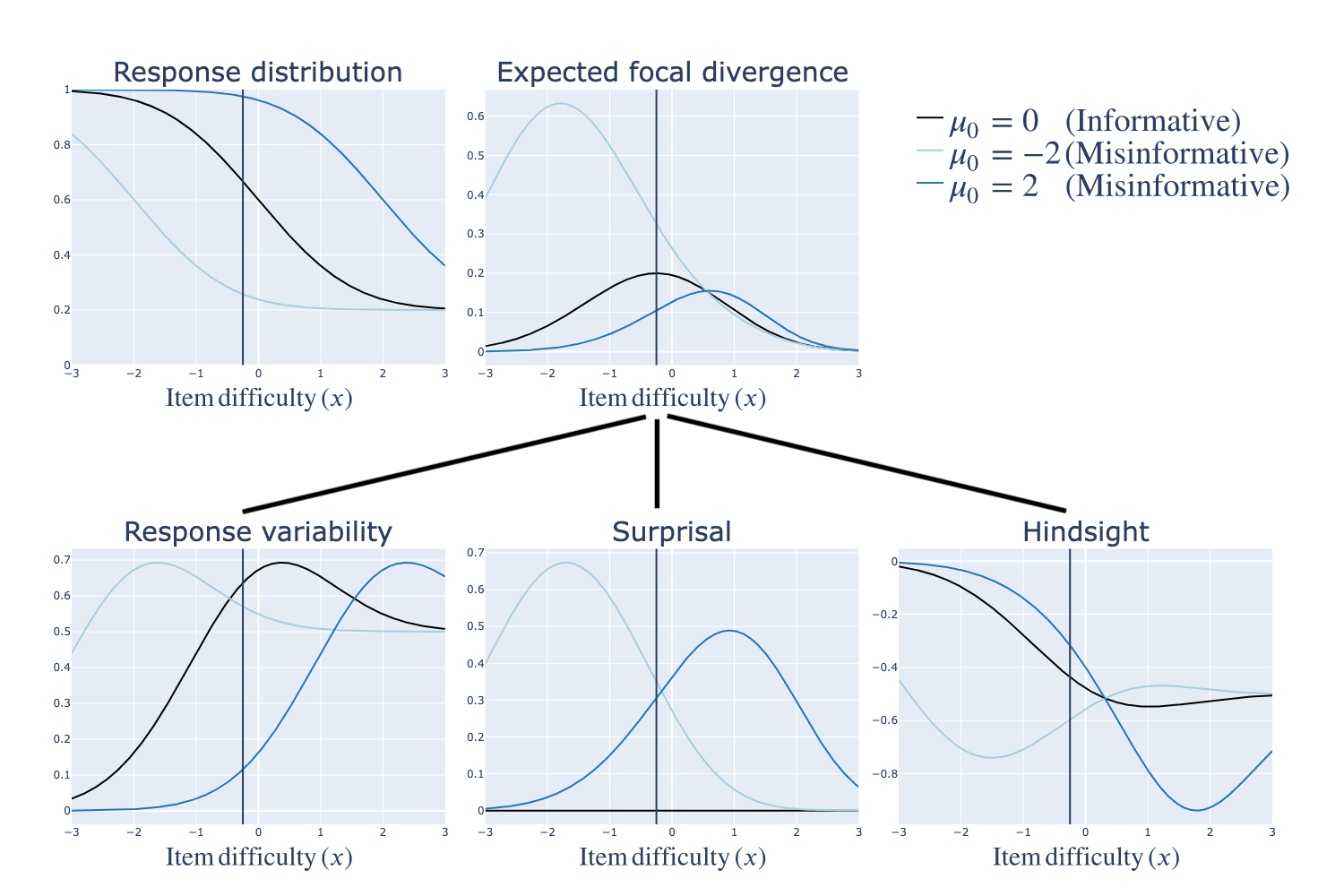}
            \notes{Figure \ref{fig:irt-motivation} reproduced along with the three components of the expected \obsU{} curves in Figure \ref{fig:irt-motivation-u1}: response variability, surprisal and hindsight (Equation \ref{eq:Ulearning}).
                As in Figure \ref{fig:irt-motivation-u1}, colors denote different true distributions $\True{\RV{\Theta}}$.
                In all cases, the prior is $\Sp{\RV{\Theta}} \sim \mathscr{N}(0, 1)$.
            }
        \end{figure}

        Figure \ref{fig:irt-learningu_priorfixed} shows the amount of response variability, surprisal and hindsight under each of the three population distributions shown in Figure \ref{fig:irt-motivation}.
        This gives insight into the puzzle posed in \S \ref{sec:motivation}: Why does the zero-centered prior lead to a more powerful experiment when the population exhibits low values of the trait $\theta$ than when the population exhibits high values of $\theta$?

        Figure \ref{fig:irt-learningu_priorfixed} reveals that this is because of two (in this case, related) reasons: Both response variability and surprisal are higher in the low-$\theta$ population.
        Looking more carefully at the response distributions shown in the lefthand panel, the probability of an observation of $y = 1$ is closest to .5 in the low-$\theta$ population.
        This makes sense: As discussed in \S \ref{sec:motivation}, low values of the trait lead to arbitrary responses --- i.e., responses that are harder to predict.
        Thus, response variability is much higher.
        For the same reason, surprisal is also higher: The low-$\theta$ population surprises the specified prior by producing $y = 0$ much more often than it anticipates. 
        (The high-$\theta$ population also surprises the specified prior by producing $y = 0$ less often than it anticipates, but the surprise is not as much as in the low-$\theta$ population.)

    This section has shown that when the specified prior is misinformed, ADO's global utility may mislead about the expected \obsU{}.
    The following sections explore the practical relevance of this misalignment.
    As stressed in \S \ref{sec:intro}, the motivation for our work is a situation where the population distribution is inaccessible to the experimenter.
    While our motivating example applied our framework to understanding the effect of variation in the population distribution, what is of more practical interest is what can be controlled by the experimenter: the prior they use, and whether they use ADO at all.
    \S \ref{sec:paramest} and \S \ref{sec:modelest} address these questions in the context of parameter estimation and model selection, respectively.

\section{Prior misinformation in the context of parameter estimation}\label{sec:paramest}
    \S \ref{sec:two-lives} and \S \ref{sec:efd} showed that under prior misinformation, ADO can be mistaken about the expected gain in information from a particular stimulus.
    In cases where it cannot reliably anticipate the expected \obsU{}, does ADO still enjoy an advantage over other experimental design methods?
    In this section, we investigate this question in the context of the problem of parameter estimation.
    We show that even under prior misinformation, ADO facilitates identification of the correct parameter value faster than other sequential design methods.
    In many practical cases, using methods like ADO may be even more important when there is danger of prior misinformation, since this misinformation can be overcome comparably faster than under other experimental design methods.

    As discussed in \S \ref{sec:uprior}, when identifying properties of specified priors that are robust to prior misinformation, we are most interested in their effect on hindsight.
    With reference to Equation \ref{eq:Ulearning}, hindsight is composed of three terms: $\pprior{\cond{y}{x,\phi}}$, $\pprior{\cond{\phi}{y,x}}$ and $\ppop{\cond{y}{x}}$.
    In the case of parameter estimation, these become $\pprior{\cond{y}{x,\theta}}$, $\pprior{\cond{\theta}{y,x}}$ and $\ppop{\cond{y}{x}}$.
    Here, unlike in the case where the focus is the model, the focal predictive distributions do not depend on the specified prior, i.e., $\pprior{\cond{y}{x,\theta}} = \ppop{\cond{y}{x,\theta}}$.
    Thus, of these three terms, only $\pprior{\cond{\theta}{y,x}} \propto \ppop{\cond{y}{x,\theta}} ~ \pprior{\theta}$, representing the specified posterior, depends on the specified prior.
    One way to achieve high hindsight given a misinformative prior is to specify a prior for which the likelihood dominates the posterior.
    As we discussed in \S \ref{sec:types-of-priors}, this is the definitional property of priors that are uninformative in parameter space.
    Indeed, empirical studies by \citet{alcala-quintana_role_2004} showed that in the context of the adaptive estimation of psychometric functions, uniform priors led to less bias than other commonly specified priors.
    These results lead us to expect that priors that are uninformative in parameter space will contribute to robustness in the face of prior misinformation.

    \subsection{Empirical results}\label{sec:param-experiments}
        This section empirically tests the robustness of ADO to misinformation in two modeling paradigms: the item response paradigm introduced in \S \ref{sec:motivation}, and a paradigm used to measure a participant's capacity for memory retention.
        All experiments reported in this paper were run using the \texttt{pyBAD} package for ADO \citep{sloman_towards_2022}.

        \subsubsection{Item response theory}\label{sec:param-experiments-irt}
            This section discusses simulation experiments to estimate the parameters of item-response models run under the modeling paradigm used as our motivating example.

            \paragraph{Experimental set-up}
                We simulated experiments under two design methods: ADO and a fixed design method.
                Again drawing inspiration from \citet{weiss_biased_1983}, who discretized the parameter space into 31 equally-spaced levels ranging from -3 to 3, the fixed design was set \emph{a priori} as all such 31 stimuli (presented in a random order).
                ADO was similarly constrained to select from amongst these 31 candidate stimuli.
                All experiments were run for 31 trials.
                For the fixed design this means that each stimulus would be been presented exactly once, while in the ADO experiments some of those candidate stimuli may be repeated or not presented at all.
                For each combination of design method, population distribution, and specified prior, we simulated a total of 1,000 experiments.
                In each experiment, a new value of $\theta^*$, the parameter value governing the true distribution of responses, was sampled at random from the corresponding population distribution, and held fixed for that experiment.
                Data were generated according to Equation \ref{eq:irt}.
                Both methods were initialized with the specified prior.
                
                We here show the results of three sets of experiments:

                \begin{enumerate}
                    \item Experiments that show the effect of \underline{changes in population distribution}, with the specified prior held fixed, were run under the same conditions as shown in Figures \ref{fig:irt-motivation} and \ref{fig:irt-learningu_priorfixed}.
                    \item Experiments that show the effect of \underline{uncontrolled changes in specified prior} fixed the population distribution to $\True{\RV{\Theta}} \sim \mathscr{N}(2, 1)$ and varied the specified prior among an informative prior ($\Sp{\RV{\Theta}} = \True{\RV{\Theta}} \sim \mathscr{N}(2, 1)$), a misinformative prior ($\Sp{\RV{\Theta}} \sim \mathscr{N}(0, 1)$), and a more dispersed misinformative prior, i.e., a prior that is uninformative in parameter space ($\Sp{\RV{\Theta}} \sim \mathscr{N}(0, 2)$).
                        We refer to these manipulations as ``uncontrolled'' changes because they do not control for the degree of prior misinformation: The uninformative prior assigns a higher prior log probability to $\theta^*$, and induces lower surprisal across part of the stimulus space.
                        Thus, the misinformative prior is at an initial disadvantage but may learn faster because of the mismatch in surprisal.
                    \item To isolate the effect of dispersion from prior misinformation, experiments that show the effect of \underline{controlled changes in specified prior} fixed the population distribution to $\True{\RV{\Theta}} \sim \mathscr{N}(0, 1)$ and varied the specified prior among an informative prior ($\Sp{\RV{\Theta}} = \True{\RV{\Theta}} \sim \mathscr{N}(0, 1)$), a misinformative prior ($\Sp{\RV{\Theta}} \sim \mathscr{N}(0, .65)$) and a more dispersed misinformative prior, i.e., a prior that is uninformative in parameter space ($\Sp{\RV{\Theta}} \sim \mathscr{N}(0, 2)$).
                        While these conditions are more artificial than those in our second set of experiments, they control for prior misinformation in the sense that the uninformative prior both tends to assign a lower prior log probability to $\theta^*$, and induces higher surprisal across the entire stimulus space.
                \end{enumerate}

\begin{figure}
                    \caption{}
                    \label{fig:irt-results}
                    {\centering
                        \begin{subfigure}{.48\linewidth}
                            \caption{}
                            \label{fig:irt-results-priorfixed}
                            \includegraphics[width=\linewidth]{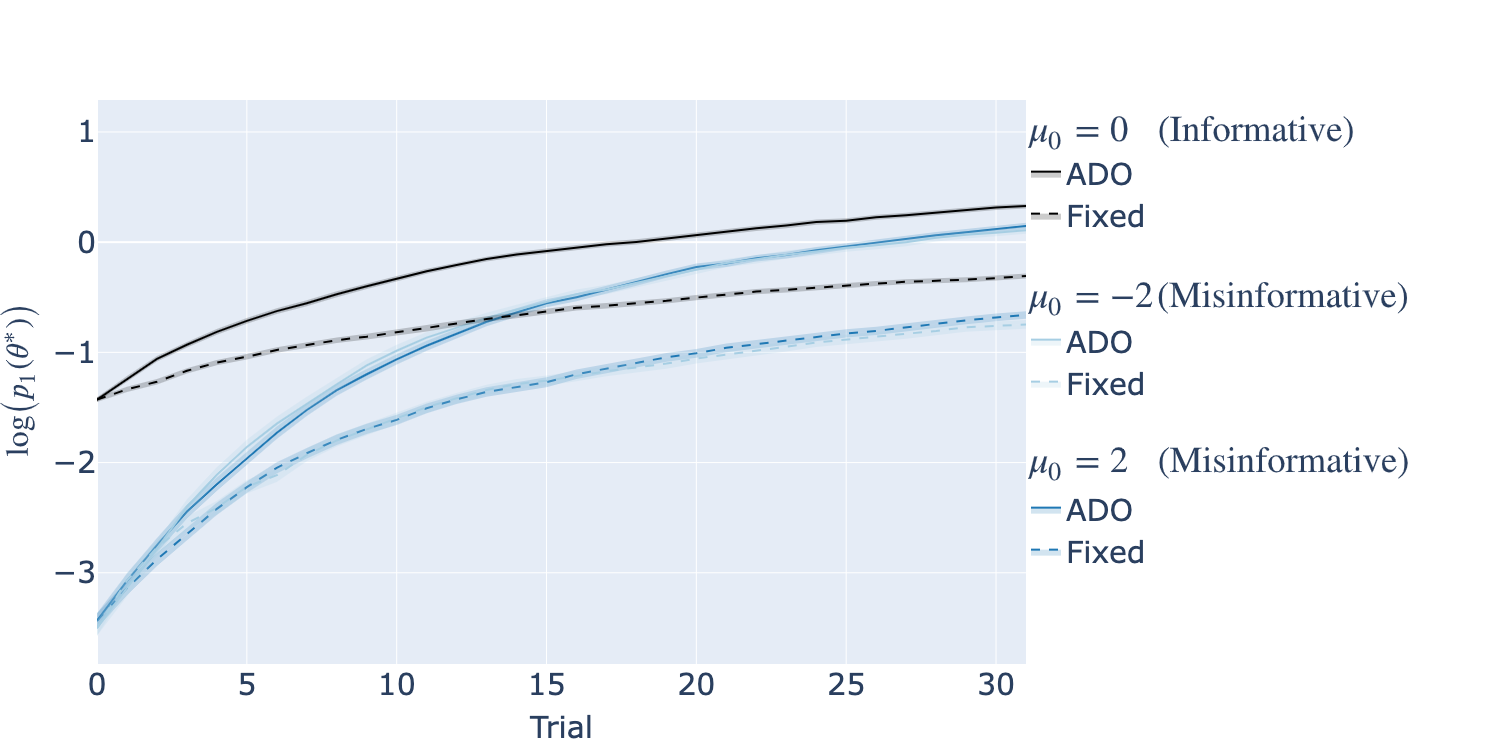}
                        \end{subfigure}
                        
                        \begin{subfigure}{.48\linewidth}
                            \caption{}
                            \label{fig:irt-results-popfixed2}
                            \includegraphics[width=\linewidth]{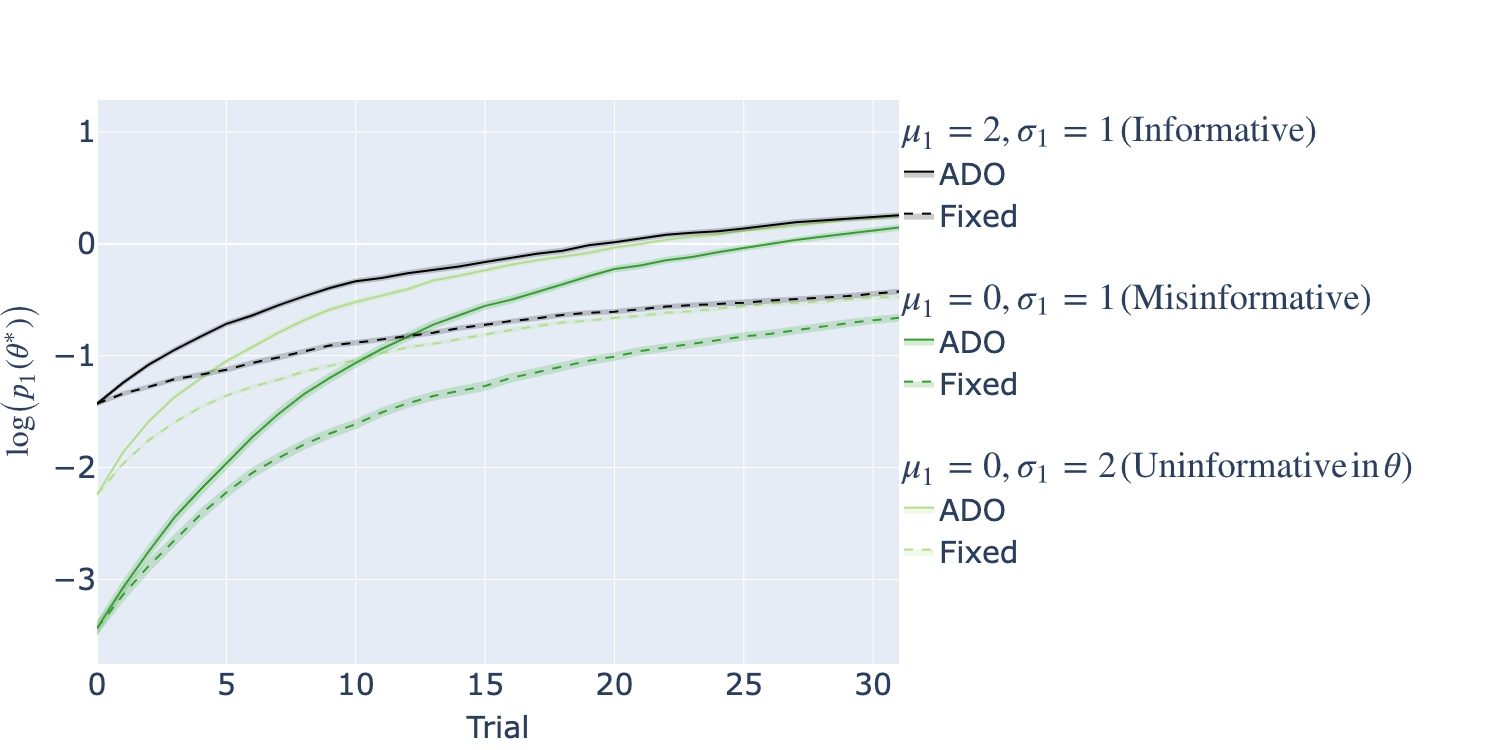}
                        \end{subfigure}\hfill\begin{subfigure}{.48\linewidth}
                            \caption{}
                            \label{fig:irt-results-popfixed0}
                            \includegraphics[width=\linewidth]{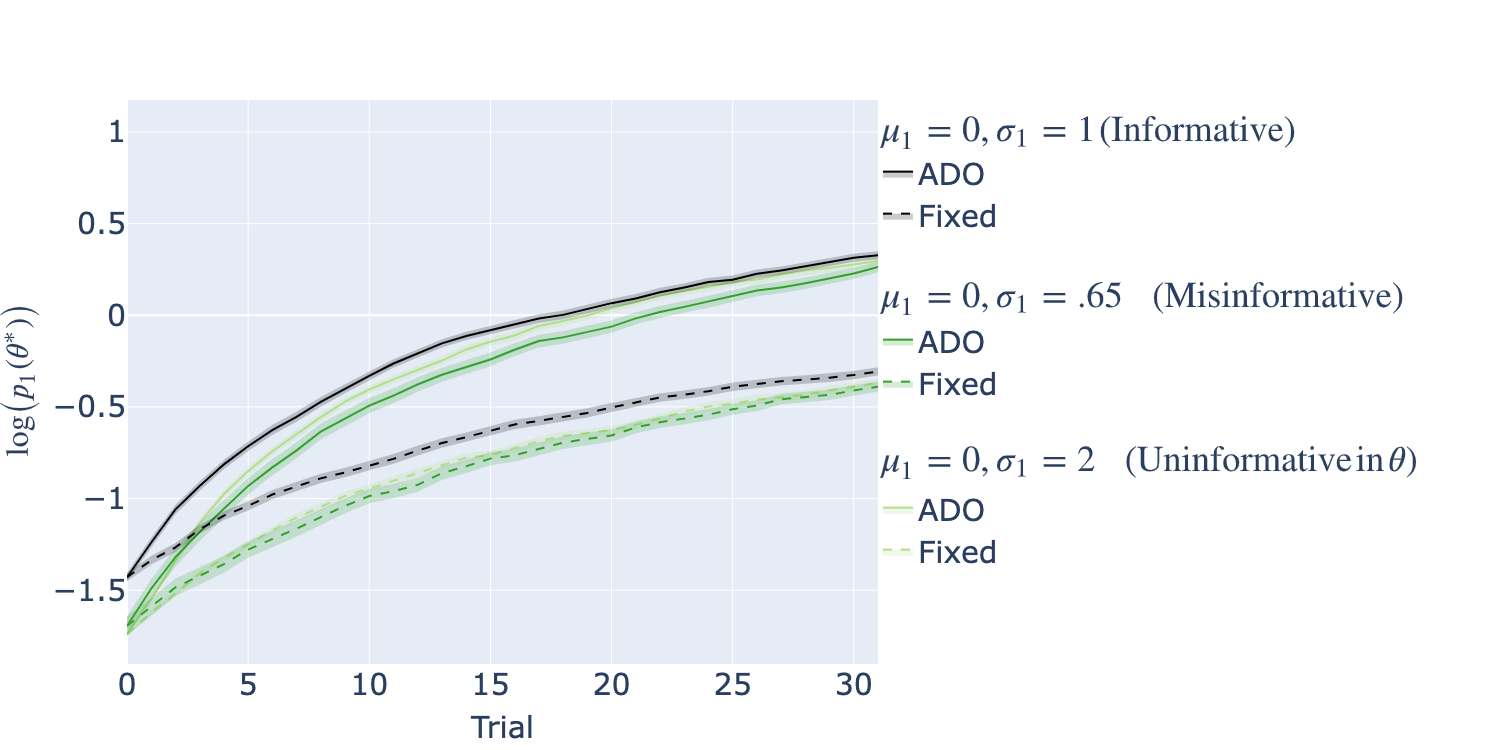}
                        \end{subfigure}
                    }
                    \notes{Item response models: Empirical results.
                      $x$-axes: Trial of the experiment.
                      $y$-axes: Log probability assigned by the specified prior to the true parameter value ($\log{\left( \pprior{\theta^*} \right)}$).
                      Lines denote means, and shaded regions denote standard errors around those means (across $n = 1,000$ simulation experiments).
                      Black curves always denote the case where the specified prior is informative, i.e., $\Sp{\RV{\Theta}} = \True{\RV{\Theta}}$.
                      Solid lines show the performance of ADO, and dashed lines show the performance of the fixed design.
                      \textbf{(a)} \textbf{Changes in population prior} (corresponds to Figure \ref{fig:irt-learningu_priorfixed}): $\Sp{\RV{\Theta}} \sim \mathscr{N}(0, 1)$. $\True{\RV{\Theta}} \sim \mathscr{N}(-2, 1)$ (light blue) vs. $\True{\RV{\Theta}} \sim \mathscr{N}(2, 1)$ (dark blue).
                      \textbf{(b)} \textbf{Uncontrolled changes in specified prior}: $\True{\RV{\Theta}} \sim \mathscr{N}(2, 1)$. $\Sp{\RV{\Theta}} \sim \mathscr{N}(0, 1)$ (dark green) vs. $\Sp{\RV{\Theta}} \sim \mathscr{N}(0, 2)$ (light green).
                      \textbf{(b)} \textbf{Controlled changes in specified prior}: $\True{\RV{\Theta}} \sim \mathscr{N}(0, 1)$. $\Sp{\RV{\Theta}} \sim \mathscr{N}(0, .65)$ (dark green) vs. $\Sp{\RV{\Theta}} \sim \mathscr{N}(0, 2)$ (light green).
                    }
                \end{figure}

            \paragraph{Results}
                Each panel of Figure \ref{fig:irt-results} shows results corresponding to one of the three sets of experiments described above.
                The $x$-axis of each panel indicates the trial number.
                The $y$-axis indicates the log posterior probability of the true parameter value.\footnote{
                    The true parameter value was different in each simulated experiment, so, writing $\theta^*_i$ for the true parameter value in experiment $i$, the average log posterior probability of the true parameter value is $\frac{\sum_{i=1}^{1000} \log \left( \pprior{\theta^*_i} \right)}{1000}$.
                }\footnote{
                    When discussing our results, we measure the effectiveness of each design method by tracking $\log \left( \pprior{\theta^*} \right)$ across trials.
                    The log transformation reflects the structure of the global utility and expected focal divergence measures.
                    Sometimes, qualitative trends of the non-logged probabilities differ from those shown in the figures in the main text.
                    For completeness, we include corresponding plots of non-logged probabilities in Appendix \ref{ap:linearp}.
                }
                In all cases, the black curve corresponds to the informative case, where the specified prior matches the population distribution.

                First, comparing ADO (solid lines) to the fixed design (dashed lines), it is clear that ADO outperforms the fixed design in all three cases.
                In fact, ADO even under prior misinformation ultimately results in stronger inference than the fixed design under an informative prior.

                Taking a closer look at the first set of simulations in Figure \ref{fig:irt-results-priorfixed}, we find no discernible difference.
                Although Figure \ref{fig:irt-learningu_priorfixed} showed the low-$\theta$ population induced higher expected \obsU{}, this difference does not translate into a difference in the rate of convergence on the correct parameter value.
                Recall from \S \ref{sec:motivation-unpacked} that the higher expected \obsU{} in the low-$\theta$ population was largely driven by higher response variability.
                If high response variability stems mainly from dispersion across values of the focus, this indicates that each value of the focus makes distinct predictions, facilitating identification of the correct value \citep{houlsby_collaborative_2011}.
                However, in the low-$\theta$ population, response variability stems mostly from higher guessing rates.
                More generally, as this example illustrates, high response variability that is inherent in the model, i.e., that does not disappear even when conditioning on a particular parameter value, inhibits identification of the correct parameter value.

                Figures \ref{fig:irt-results-popfixed2} and \ref{fig:irt-results-popfixed0} show that the prior that is uninformative in parameter space generally converges more quickly on the correct parameter value, whether using ADO or the fixed design.\footnote{
                    The difference appears small in the log space, but is illustrated more dramatically when probabilities are plotted on the linear scale, as shown in Figure \ref{fig:irt-results-popfixed0-linearp}.
                }
                This is the case even when controlling for prior misinformation (Figure \ref{fig:irt-results-popfixed0}), since priors that are uninformative in parameter space are able to respond more effectively to unexpected observations.

        \subsubsection{Memory retention}\label{sec:param-experiments-memreten}
            While the simplicity of the item-response paradigm allows careful control of our experimental conditions and facilitates interpretation, it potentially limits the generalizability of our findings.
            We now test whether the main finding --- that ADO for parameter estimation outperforms other sequential design methods under prior misinformation --- holds in a more complex modeling paradigm: estimating a participant's capacity for memory retention.
            
            Over a century of research on forgetting has shown that a person’s ability to remember information just learned drops quickly for a short time after learning and then levels off as more and more time elapses \citep{ebbinghaus_memory_1913,laming_analysis_1992}. 
            The simplicity of this data pattern has led to the introduction of a number of models to describe the rate at which information is retained in memory \citep{rubin_one_1999}.  
            One of these is the power-law model, which posits that the probability a participant will recall an item ($y = 1$) $x$ seconds after presentation is \citep{wixted_form_1991}:
            \begin{align}\label{eq:pow}
                p(y = 1) &= a(x + 1)^{-b}.
            \end{align}

            The parameters of the model are $a$ and $b$, where $0 \le a \le 1$ encodes a baseline level of accuracy, and $0 \le b \le 1$ encodes the forgetting rate.

            \paragraph{Experimental set-up}
                We again ran experiments under two design methods: ADO and a fixed design method.
                The design variable to be manipulated was the time delay between presentation of the target and the recall phase (i.e., $x$ in Equation \ref{eq:pow}).
                The fixed design method was a slight variation on a benchmark used by \citet{cavagnaro_adaptive_2010}, taken from previous literature \citep{rubin_precise_1999}.
                In the fixed design method scheme, delays were $\{ 0, 1, 2, 4, 7, 12, 21, 35, 59, 99 \}$.
                Each fixed-design experiment ran for 100 trials, allowing each of these 10 delays to be repeated 10 times.
                The order of stimuli was randomized separately for each experiment.
                ADO experiments also ran for 100 trials.
                In each ADO trial, the time delay could be any integer between 0 and 100 seconds.

                We simulated experiments under two different population distributions, each combined with four types of specified priors.
                For the \underline{high $b$} population, we set $\boldsymbol{b}_0 \sim \mathrm{Beta}(2,1)$, i.e., the forgetting rate is high, on average, but negatively skewed.
                For the \underline{low $b$} population, we set $\boldsymbol{b}_0 \sim \mathrm{Beta}(1,2)$, i.e., the forgetting rate is low, on average, but positively skewed.
                For both populations, we set $a$ to $\mathrm{Beta}(1, 1)$, which is equivalent to a uniform distribution between 0 and 1.
                The four types of specified priors are as follows:

                \begin{enumerate}
                    \item \underline{Informative} priors matched the population distributions given above.
                    \item Priors that mistook the two populations: The specified prior for the high $b$ population was $a \sim \mathrm{Beta}(1,1), b \sim \mathrm{Beta}(1,2)$, and the specified prior for the low $b$ population was $a \sim \mathrm{Beta}(1,1), b \sim \mathrm{Beta}(2,1)$.
                        In the context of these experiments, we refer to these as \underline{misinformative} priors.
                    \item Priors that were \underline{uninformative in parameter space} specified that $a \sim \mathrm{Beta}(1,1), b \sim \mathrm{Beta}(1,1)$.
                    \item Priors that were \underline{uninformative in data space} resulted in maximally dispersed predictive distributions.
                        The prior that achieves this is $a \sim \mathrm{Beta}(2,1), b \sim \mathrm{Beta}(1,4)$ \citep{cavagnaro_adaptive_2010}.\footnote{
                            Note that it is only when the prior is uninformative in data space that the distribution over $a$ is misspecified.
                        }
                \end{enumerate}

\begin{figure}
                    \caption{}
                    \label{fig:memreten-paramest}
                    \begin{subfigure}{.48\linewidth}
                        \caption{}
                        \label{fig:memreten-paramest-pred}
                        \includegraphics[width=\linewidth]{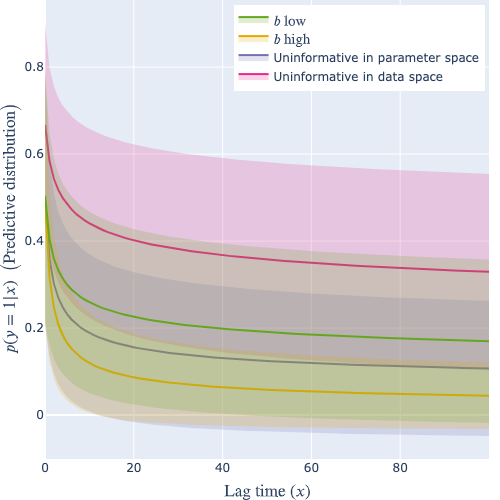}
                    \end{subfigure}\hfill\begin{subfigure}{.48\linewidth}
                        \caption{}
                        \label{fig:memreten-paramest-absolute}
                        \includegraphics[width=\linewidth]{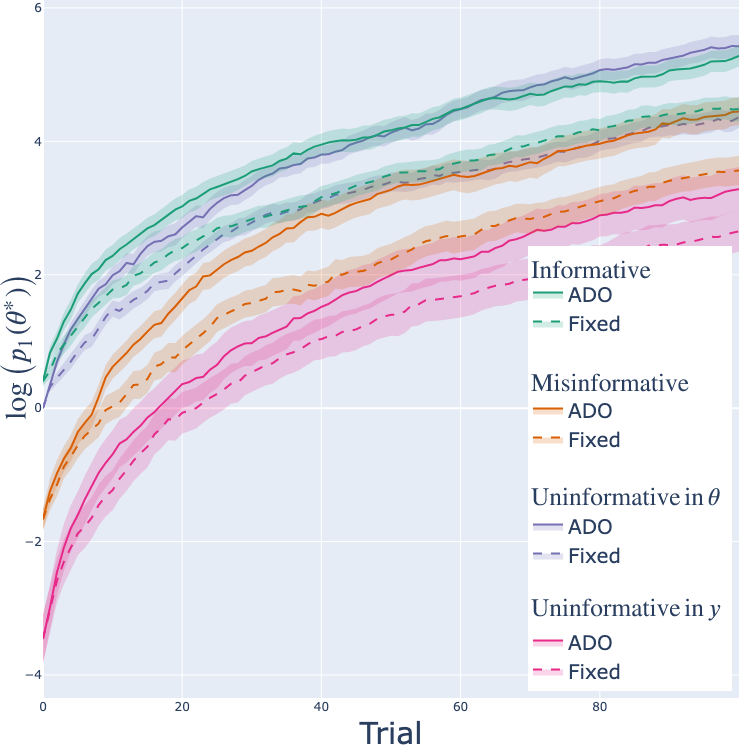}
                    \end{subfigure}
                    \notes{Memory retention models (parameter estimation): Empirical results.
                        \textbf{(a)} Predictive distributions.
                        Lines denote mean predictions, and shaded regions denote the standard deviation across the specified prior.
                        \textbf{(b)} Performance across the course of the experiment.
                            $x$-axis: Trial number.
                            $y$-axis: $\log{\left( \pprior{\theta^*} \right)}$.
                            Lines denote means, and shaded regions denote standard errors around those means (across $n =$ 2 populations $\times$ 100 repetitions = 200 simulated experiments).
                    }
                \end{figure}

                Figure \ref{fig:memreten-paramest-pred} shows typical forgetting curves under each prior.

                For each population and for each type of prior, we simulated 100 experiments, for a total of 2 design methods $\times$ 2 populations $\times$ 4 types of specified priors $\times$ 100 repetitions = 1,600 experiments.
                In each experiment, a true parameter $\theta^*=\{a^*,b^*\}$ was randomly drawn from the corresponding population distribution, the time delay on each trial was selected according to the design method, and data were generated according to Equation \ref{eq:pow}.

            \paragraph{Results}
                Figure \ref{fig:memreten-paramest-absolute} shows how the correctness of inference evolves over the course of the experiment under each type of prior (results are pooled across the two populations).
                Values on the $y$-axes are the log probabilities assigned to the true, generating parameter value under each specified prior.
                This figure shows replication of our main result from \S \ref{sec:param-experiments-irt}: ADO outperforms the benchmark for each population and every type of specified prior.
                Interestingly, unlike in the item-response paradigm, differences in performance at the end of the experiment are mostly accounted for by the type of prior: The fixed design under the informative prior generally does better than ADO under the misinformative or uninformative in data space priors (this is despite the fact that, unlike in the item-response example, ADO has access to a larger stimulus bank than the fixed design method).
            
        In sum, in both simulation paradigms, ADO performed better than the fixed design method even under prior misinformation.
        In other words, we do not find that prior misinformation diminishes ADO's relative advantage.
        In fact, our results suggest that using ADO when there is prior misinformation may help to overcome that misinformation more quickly than using other design methods.

\section{Prior misinformation in the context of model selection}\label{sec:modelest}
    \S \ref{sec:paramest} showed that in the context of parameter estimation, ADO usually leads to faster convergence on the true parameter value under prior misinformation than other sequential design methods.
    This section explores whether the same can be said in the context of model selection.
    It will turn out that, in the context of model selection, the effect of prior misinformation can be more damaging: It can lead one to favor the wrong model.

    A common measure of the strength of evidence in favor of one model $m_1$ over another model $m_2$ is the Bayes factor, or relative likelihood of data $\cond{y}{x}$ under $m_1$ and $m_2$:
    \begin{align}\label{eq:bf}
        BF(m_1,m_2) &= \frac{p_1(\cond{y}{x,m_1})}{p_1(\cond{y}{x,m_2})} \nonumber \\
        &= \frac{\int_\theta p(\cond{y}{x,\theta}) ~ p_1(\cond{\theta}{m_1})}{\int_\theta p(\cond{y}{x,\theta}) ~ p_1(\cond{\theta}{m_2})}.
    \end{align}

    Equation \ref{eq:bf} reveals the sensitivity of model selections to prior misinformation: The apparent strength of evidence in favor of one model over the other is a function of the specified priors $\cond{\Sp{\RV{\Theta}}}{m_1}$ and $\cond{\Sp{\RV{\Theta}}}{m_2}$.
    Under prior misinformation, the magnitude and even direction of the Bayes factor can be misleading --- implying that it can lead to the erroneous selection of one model over the true, generating model \citep{vanpaemel_prior_2010,lopes_confronting_2011}.
    
    This is an important concern in Bayesian inference, and addressing it through the choice of prior has been the subject of much literature \citep{vanpaemel_prior_2010,lee_robust_2019}.
    In this section, we show that this relates importantly to the consequences of the choice of prior in its decision-theoretic role.

    Recall Equation \ref{eq:Umodel}, which gives the global mutual information utility in the context of model selection.
    \citet{cavagnaro_adaptive_2010} showed that Equation \ref{eq:Umodel} can be rewritten as a function of the Bayes factors between all pairs of candidate models.
    This result implies that ADO results in the selection of stimuli that are expected to lead to extreme Bayes factors according to the specified prior.
    When the Bayes factors are misleading, this effect of ADO can exacerbate the amount of information encountered that leads one to the wrong model.

    The results presented in the remainder of this section will show that in the case of model selection, like in the case of parameter estimation, ADO tends to accelerate convergence towards a particular model.
    However, under a deceptive prior, this might be the wrong model.
    In such cases, desirable behavior for an experimental design method would be to decelerate, rather than accelerate, convergence.
    We will show in \S \ref{sec:model-experiments} that in such cases other experimental design methods outperform ADO.

    \subsection{Effect of prior misinformation through the lens of Bayesian inference}\label{sec:bi-model}
\begin{figure}
            \caption{}
            \label{fig:model-ex}
            \begin{subfigure}{.48\linewidth}
                \caption{}
                \label{fig:model-ex-priors}
                \includegraphics[width=\linewidth]{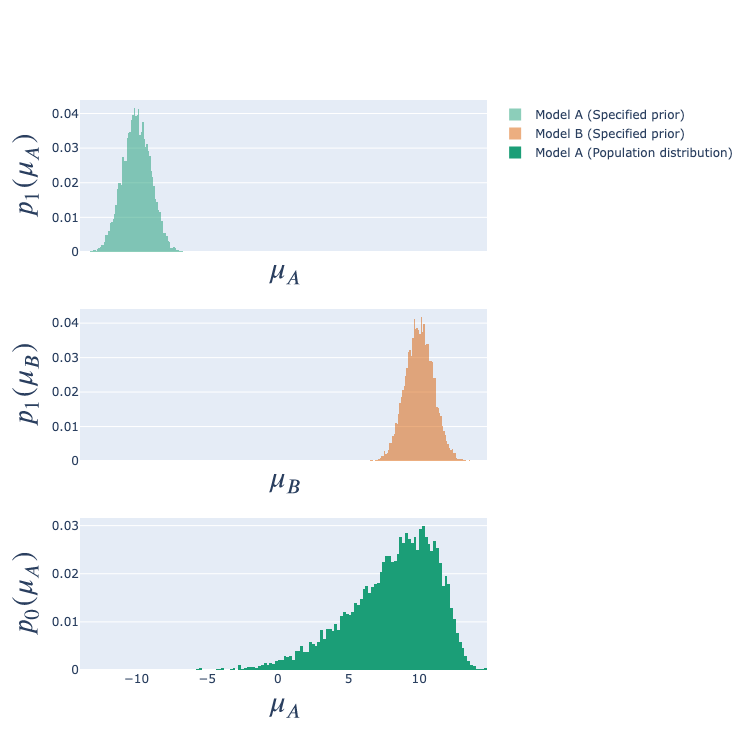}
            \end{subfigure}\begin{subfigure}{.48\linewidth}
                \caption{}
                \label{fig:model-ex-preds}
                \includegraphics[width=\linewidth]{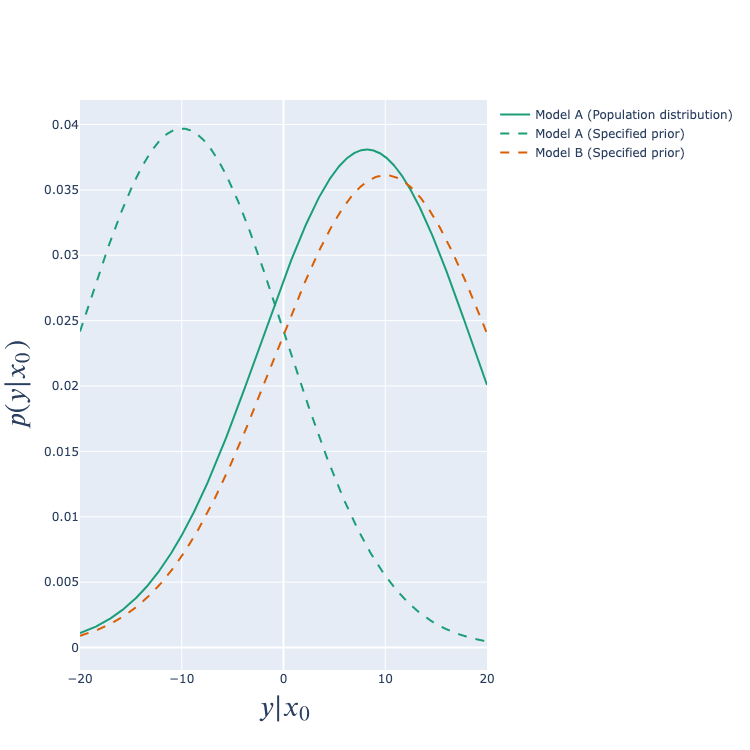}
            \end{subfigure}
            \notes{Prior misinformation biases inference for model selection: Motivating example (see discussion in main text).
                \textbf{(a)} Parameter distributions.
                \textbf{(b)} Focal predictive distributions.
            }
        \end{figure}

        Before turning to our results on the effect of ADO, we first present a simple example illustrating the potential effect of prior misinformation in the context of Bayesian inference more generally.
        Consider the toy example shown in Figure \ref{fig:model-ex}.
        The task is to distinguish between two models, Model A and Model B.  
        Each model has a free parameter, $\mu_A \sim \boldsymbol{\mu}_A$ and $\mu_B \sim \boldsymbol{\mu}_B$, respectively,\footnote{
            Here we simply bold the notation for the realization $\mu_A$ ($\mu_B$) to indicate its corresponding distribution, $\boldsymbol{\mu}_A$ ($\boldsymbol{\mu}_B$), to avoid confusion with the random variable that generically represents the distribution over models, $\RV{M}$.
        } and makes predictions for a single stimulus $x_0$.  
        Under Model A, responses to $x_0$ are distributed as $\cond{\RV{Y}}{x_0, \mu_A} \sim \mathscr{N}(\mu_A, 10)$, while under Model B, responses to $x_0$ are distributed as $\cond{\RV{Y}}{x_0, \mu_B} \sim \mathscr{N}(\mu_B, 11)$.  
        Thus, the families of functions captured by the two models are distinguished by the inherent variance in responses. 
        The experimenter is required in advance of the experiment to assign a prior distribution to $\RV{\left( M,\boldsymbol{\mu}_A,\boldsymbol{\mu}_B \right)}$, i.e., to both assess the relative likelihoods of Model A and Model B and to specify the distributions over $\boldsymbol{\mu}_A$ and $\boldsymbol{\mu}_B$.

        The top two panels of Figure \ref{fig:model-ex-priors} show the prior parameter distributions the experimenter specifies for Models A and B.
        The corresponding focal predictive distributions for the two models are shown, respectively, as the green and orange dashed lines in Figure \ref{fig:model-ex-preds}.

        Now, consider a heterogeneous participant population in which everyone responds according to Model A (i.e., $\sigma = 10$), but with different values of $\mu_A$ as represented in the bottom panel of Figure \ref{fig:model-ex-priors}.
        The solid green line in Figure \ref{fig:model-ex-preds} shows the distribution of responses from this population.
        The dispersion in this curve captures both the inherent variance in each participant's responses ($\sigma = 10$), and variance due to the distribution of values of $\mu_A$ across participants.
        Importantly, most participants will produce data that is more likely under Model B than under Model A, under their respective specified priors, yielding apparently strong evidence in favor of Model B.

        The upshot is that the true state of the world, i.e., the true response distribution, may look very different from the focal predictive distribution corresponding to the generating model.
        In essence, the specified prior sets an expectation for what data from a given model will look like, but data from that model may look different in reality if the specified prior is far from the population distribution, and that can lead to wrong inference.
        In effect, unless the true state of the world happens to coincide exactly with the predictive distribution of $m^*$, each possible value of the focus is effectively misspecified \emph{a priori}.
        Notice that this doesn't matter in the case of parameter estimation: In this case, the focal predictive distributions are unaffected by prior misinformation --- as Equation \ref{eq:predictive} shows, they are a function only of the model structure, which is (by assumption) known.

        This example is albeit quite contrived to prove a point.
        However, such deceptive priors --- priors that induce initial convergence towards the wrong model --- can actually emerge in practice, as we show in \S \ref{sec:model-experiments}.
        In the remainder of this section, we explore --- conceptually in \S \ref{sec:bdt-modelest} and empirically in \S \ref{sec:model-experiments} --- the degree to which this phenomenon persists in the context of ADO.
        The consistency of Bayesian inference guarantees that the experimenter in this example will eventually be able to recover $m^*$.
        However, when the amount of data collected is not large, relying on Bayesian decision-theoretic policies --- i.e., choosing data on the basis of these misinformed inferences --- has the potential to exacerbate the effect of misinformation.

    \subsection{Effect of prior misinformation through the lens of Bayesian decision theory}\label{sec:bdt-modelest}
        In the toy example in \S \ref{sec:bi-model}, ADO would assign $x_0$ a high global utility because it induces a large divergence between the predictions of Models A and B --- even though these predictions are made on the basis of prior misinformation.
        
        In general, when crafting a policy for selecting optimal designs, the goals of parameter estimation and model selection may come into conflict. 
        A stimulus that ADO calculates is optimal for discriminating between models may not be optimal for refining estimates of the distribution of parameter values.
        In other words, ADO for model selection faces a version of an explore--exploit dilemma: By acting on its prior beliefs about each model's predictions, it may fail to explore parts of the sample space that could challenge these beliefs.

        Thus, when the goals of model selection and parameter estimation are in conflict, ADO can actually exacerbate the problem.
        By aggressively ``exploiting'' areas of the design space that appear to yield information about the models, ADO finds powerful evidence in favor of its prior beliefs.
        In contrast, by ``exploring'' less apparently informative stimuli, other methods may have more of an opportunity to learn the correct parameter distributions before making strong conclusions about the generating model.

        ADO's aggressiveness is thus a double-edged sword: It converges quickly on conclusions based on what it believes about the predictions of the foci.
        However, in the case where prior beliefs do not reflect the population distribution, it does not seek opportunities to challenge these incorrect beliefs.

    \subsection{Choosing a prior distribution}\label{sec:model-prior}
        \S \ref{sec:paramest} showed that, in the case of parameter estimation, priors that are uninformative in parameter space can somewhat mitigate the damage of prior misinformation.
        One would hope that the issues that arise in model selection could be avoided by using similarly uninformative priors.

        Unfortunately, this is not the case: As will be shown in the following section, priors that are uninformative in parameter space nevertheless associate models with particular response distributions, and are also prone to inducing biased inference.
        One could nevertheless hope that specifying such priors over the parameter distributions of candidate models might mitigate the problem by facilitating more rapid convergence on informative parameter distributions.
        Indeed, we find empirically that in one model selection context, recovery from biased inference is relatively fast under a uniform prior.
        However, it is difficult to disentangle the effect of the responsiveness of the uniform prior from its effect on the focal predictive distributions --- in particular, how they diverge from the response distribution.
        We leave investigating whether specifying priors that are uninformative in parameter space mitigates biased inference in the context of model selection as an avenue for future work.

        Is it possible to identify a prior that is instead ``uninformative in model space''?
        In the case of parameter estimation, the important characteristic of an uninformative prior was that it was responsive: Areas of the parameter space quickly became represented in proportion to the relative likelihood they assigned to the history of observations.
        A prior that was uninformative in model space would facilitate the proportional representation of models according to their relative conditional likelihood.
        But as emphasized in \S \ref{sec:mi-modelest}, the relative conditional likelihood of a model depends on the prior parameter distribution; indeed, the problem of not knowing the parameter distribution is in a sense the problem of not knowing the conditional likelihood distribution $\cond{\True{\RV{Y}}}{x, m}$.

        In summary, these results suggest the absence of concrete guidance for the case of model selection.
        The following section reinforces through simulation results that apparently uninformative priors can inadvertently induce biased inference.

    \subsection{Empirical results}\label{sec:model-experiments}
        This section extends the memory retention paradigm introduced in \S \ref{sec:param-experiments-memreten} to model selection.
        The goal of these results will be to demonstrate that apparently uninformative priors can inadvertently bias inference, and that this bias is exacerbated by ADO.

        In these experiments, the goal is to distinguish the power-law model introduced in \S \ref{sec:param-experiments-memreten} (Equation \ref{eq:pow}) from the exponential model of memory retention, which posits that the probability a participant will recall an item $x$ seconds after presentation is:

        \begin{align}
            p(y = 1) &= ae^{-bx}.
        \end{align}

        \paragraph{Experimental set-up}
            We considered two types of prior distributions:

            \begin{enumerate}
                \item A prior that is \underline{uninformative in parameter space} assumes that both models have prior distributions $a~\sim~\mathrm{Beta}(1, 1)$ and $b~\sim~\mathrm{Beta}(1, 1)$.
                \item A prior that is \underline{uninformative in data space} assumes that the power-law model has prior distribution $a~\sim~\mathrm{Beta}(2, 1)$ and $b~\sim~\mathrm{Beta}(1, 4)$ and that the exponential model has prior distribution $a~\sim~\mathrm{Beta}(2, 1)$ and $b~\sim~\mathrm{Beta}(1, 80)$.
                  These priors result in maximally diffuse predictive distributions for each model \citep{cavagnaro_adaptive_2010}.
            \end{enumerate}

\begin{figure}
                \caption{}
                \label{fig:memreten-model}
                \begin{subfigure}{.32\linewidth}
                    \caption{}
                    \label{fig:memreten-model-pred}
                    \includegraphics[width=\linewidth]{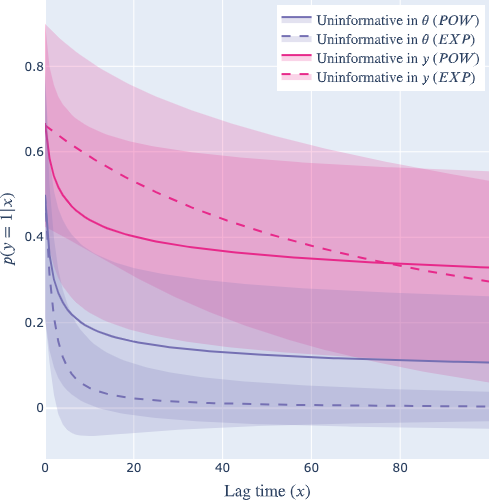}
                \end{subfigure}\hfill\begin{subfigure}{.32\linewidth}
                    \caption{}
                    \label{fig:memreten-model-popunif}
                    \includegraphics[width=\linewidth]{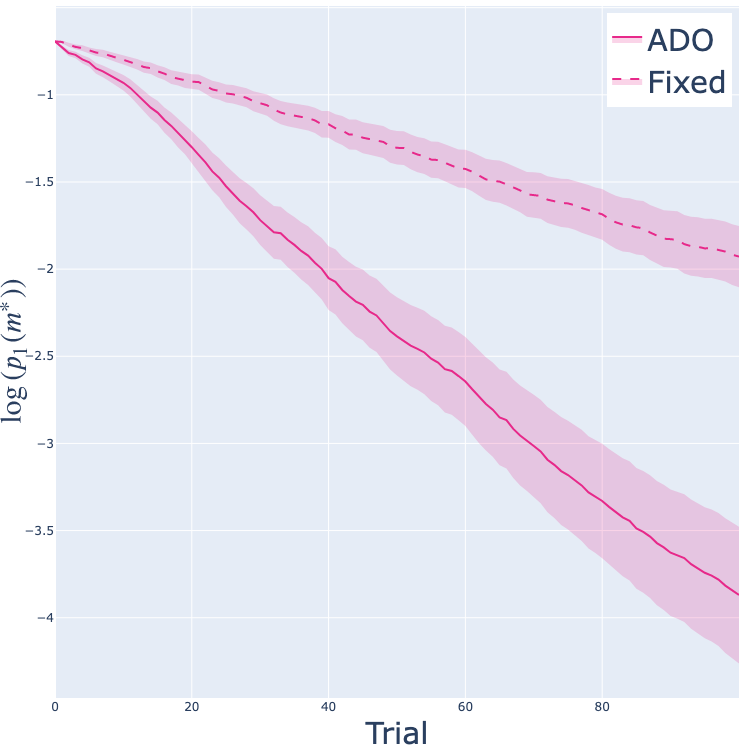}
                \end{subfigure}\hfill\begin{subfigure}{.32\linewidth}
                    \caption{}
                    \label{fig:memreten-model-popcmpk}
                    \includegraphics[width=\linewidth]{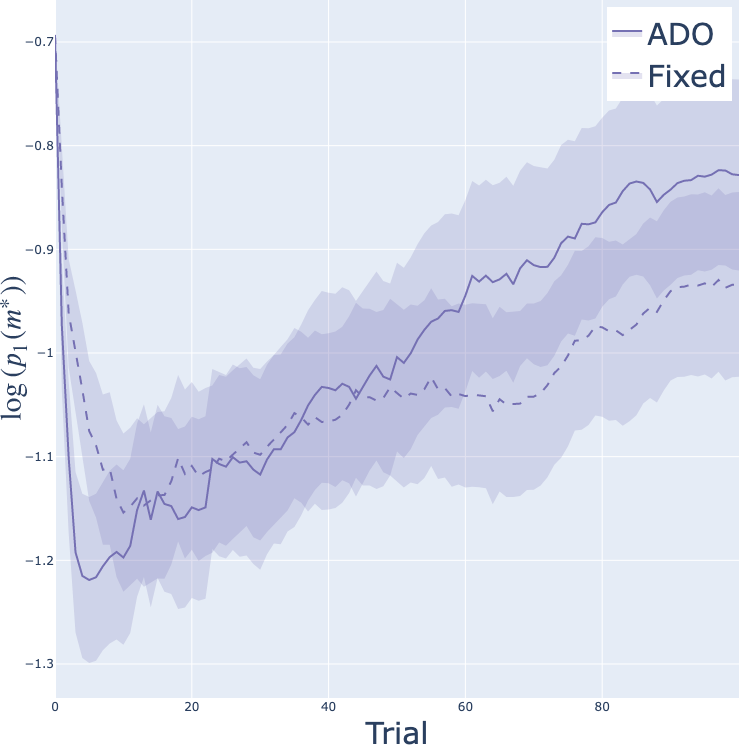}
                \end{subfigure}
                \notes{Memory retention models (model selection): Empirical results.
                  \textbf{(a)} Predictive distributions.
                  Lines denote mean predictions, and shaded regions denote the standard deviation across the specified prior.
                  \textbf{(b--c)} Absolute performance across the course of the experiment.
                  $x$-axes: Trial number.
                  $y$-axes: $\log{\left( \pprior{m^*} \right)}$.
                  Lines denote means, and shaded regions denote standard errors around those means (across $n =$ 2 models $\times$ 100 repetitions = 200 simulated experiments).
                  In Panel \textbf{b}, $\cond{\True{\RV{\Theta}}}{m}$ is uninformative in parameter space; $\cond{\Sp{\RV{\Theta}}}{m}$ is uninformative in data space, for all $m$.
                  In Panel \textbf{c}, $\cond{\True{\RV{\Theta}}}{m}$ is uninformative in data space; $\cond{\Sp{\RV{\Theta}}}{m}$ is uninformative in parameter space, for all $m$.
            }
            \end{figure}

            The predictive distributions associated with both types of priors are shown in Figure \ref{fig:memreten-model-pred}.

            We ran two sets of experiments: In one set of experiments, we sampled responses from a population distribution that matches the prior that is uninformative in parameter space, while the specified prior was instead uninformative in data space.
            In the other set, we sampled responses from a population distribution that matches the prior that is uninformative in data space, while the specified prior was instead uninformative in parameter space.
            Thus, all experiments were characterized by prior misinformation.

            Within each set, in half of the experiments, data were generated from the power-law model, while in the other half, data were generated from the exponential model (the prior over models was always correctly specified as assigning a probability of .5 to each model).
            For each set and generating model, we ran 100 experiments in which a parameter was randomly drawn from the corresponding population distribution, for a total of 2 design methods $\times$ 2 types of priors $\times$ 2 models $\times$ 100 repetitions = 800 experiments.

        \paragraph{Results}
            Figures \ref{fig:memreten-model-popunif} and \ref{fig:memreten-model-popcmpk} show how the correctness of inference evolves over the course of the experiment under each type of prior (results are pooled across the two generating models).
            Values on the $y$-axes are the log probabilities assigned to the generating model $m^*$ under each generating prior.

            Figures \ref{fig:memreten-model-popunif} and \ref{fig:memreten-model-popcmpk} exhibit the dynamic explained in the previous subsections: Inference favors the wrong model (at least initially), and this is exacerbated by ADO.
            The reasons for this are precisely the reasons for the confusion illustrated in Figure \ref{fig:model-ex}: As shown in Figure \ref{fig:memreten-model-pred}, in both cases the specified prior distributions are wildly off base about the expected behavior of the population characterized by each model.

            Recovery from biased inference under the specified prior that is uninformative in parameter space (Figure \ref{fig:memreten-model-popcmpk}) is quicker than recovery from biased inference under the specified prior that is uninformative in data space (Figure \ref{fig:memreten-model-popunif}), potentially reflecting the capacity of the prior that is uninformative in parameter space to more quickly ``respond'' to unexpected observations.
            However, our setup here is not adequate to confirm this.
            Notice first that while the specified prior varies between the two panels of Figure \ref{fig:memreten-model}, so does the population distribution.
            More fundamentally, the specified prior changes the focal predictive distributions.
            Taken together, this implies that our setup does not (and perhaps cannot) control for qualitative differences in the divergence between the focal predictive distributions and the response distribution, which, as discussed in \S \ref{sec:bi-model}, is the source of the biased inferences.
            
\section{Robust practices for ADO for model selection}\label{sec:robust-practices}
    The results from the previous section highlight the importance of taking steps to ensure one's priors are informative --- especially when used in conjunction with decision-theoretic methods like ADO, which amplify biases induced by prior misinformation.
    In the context of model selection, if an experimenter specifies a prior that faithfully captures their epistemic uncertainty, ADO will treat that uninformative prior as being a true representation of relative likelihoods in the world and select designs accordingly.
    Because the two roles of the prior here conflict, this can result in incorrect inferences.

    While we framed our results in \S \ref{sec:modelest} as applying to the problem of model selection --- identification of model structure --- note that these results apply to any situation in which knowing the value of the focus of interest does not completely identify the true data-generating distribution.
    In the case of model selection, this applies because one needs the value of both $m$ and $\theta$ to identify the data-generating distribution, yet evaluates performance based only on $m$.
    However, one could also apply ADO to, e.g., a parameter estimation problem for which some ``nuisance parameters'' are not considered foci for inference (e.g., estimating only main effects in the presence of fixed or participant-level effects).
    In these cases, our results on model selection, not parameter estimation, would apply.

    \S \ref{sec:modelest} discussed the potential beneficial effect of specifying priors that are uninformative in parameter space in mitigating these biases.
    This section discusses additional methods to alleviate or anticipate this bias, some of which have been adopted by previous studies, and some of which provide promising avenues for future research.

    \subsection{Additional trials to inform specified priors}
        One way to increase confidence in one's specified priors is to devote a portion of one's experimental resources to collecting observations from which to learn more informed parameter distributions.
        For example, when using ADO to distinguish between competing models of intertemporal choice, \citet{cavagnaro_functional_2016} devoted three quarters of each experiment to parameter estimation, i.e., selecting stimuli to maximize the global utility function for parameter estimation, before using the inferred posteriors for each participant during the later model selection trials.
            
        In a parameter estimation application, \citet{kim_hierarchical_2014} leveraged hierarchical modeling techniques to pool information across participants to construct informed distributions: Data from each sequential participant was used to refine the specified prior for the next participant.
        They showed that this method led to better parameter estimates in the context of a psychophysical experiment.

        While these methods offer promising solutions for many use cases, their application falls outside the scope considered by our work.
        As we stated in \S \ref{sec:intro}, we consider situations in which the experimenter wishes to use the same prior for every participant.
        This characterizes situations in which incorporating data from previous participants would be infeasible or unfair (e.g., educational testing), or when the experimenter cannot afford to spend scarce resources on additional parameter estimation trials.
        (Note that participants in \citet{cavagnaro_functional_2016}'s study were required to complete 80 experimental trials.
        Conducting an experiment of this length would be at best difficult and at worst impossible in cases in which candidate stimuli correspond to potentially irritating or invasive tests such as a medical procedure.)

    \subsection{Total entropy utility}\label{sec:total-ent}
        \citet{borth_total_1975} introduced the total entropy utility function in order to cope with the dual sources of uncertainty that characterize the model selection problem, i.e., uncertainty about both the model identifier and the parameter value.
        The total entropy utility function considers the entire state of the world as the focus of the utility function:

        \begin{align}\label{eq:Utotalent}
            U(x) &= \sum_{m \in M} p(m) \int_\theta \int_y \log{\left( \frac{p(\cond{m,\theta}{y,x})}{p(m,\theta)} \right)} ~ p(\cond{y}{x,\theta,m}) ~ p(\cond{\theta}{m}).
        \end{align}

        We had hoped that running ADO using the total entropy utility function would, like \citet{cavagnaro_functional_2016}'s method, lead to a balance between parameter estimation and model selection trials.
        We had further hoped that it would do so more efficiently than fixed or heuristic methods of achieving this balance.
            
        To test this, we ran simulation experiments with exactly the same setup as those discussed in \S \ref{sec:model-experiments}, with the exception that when using ADO, the stimulus that maximized Equation \ref{eq:Utotalent} (rather than Equation \ref{eq:Umodel}) was selected.
        The results of these experiments, presented in Appendix \ref{ap:total-entropy}, did not show a consistent advantage of the total entropy utility function in leading to more robust selection of the correct model.

    \subsection{Novel approaches to robust adaptive experiments}\label{sec:novel-approaches}
        The previous two subsections discussed existing methods for coping with the effect of prior misinformation on model selection.
        However, these existing methods can be prohibitively costly (running additional trials to inform priors) or potentially ineffective (using the total entropy utility function).
        An important direction for future research is the development of methods that increase the robustness of adaptive design methods to the pitfalls introduced in \S \ref{sec:modelest}.
        To this end, in this section, we propose two steps experimenters can take in the design and implementation of adaptive experiments to increase their robustness.
        We leave further development and stress testing of these approaches as avenues for future research.

        \paragraph{1. Anticipating biases via prior sensitivity analyses}
            As mentioned in \S \ref{sec:two-lives}, the choice of prior distribution in the context of Bayesian inference is the topic of a substantial literature.
            One practice advocated in this literature (e.g., \citet{lee_robust_2019}) is to perform prior sensitivity analyses, i.e., to perform data analysis under a variety of priors to ensure one's inferences are robust to the specification of the prior.
            
            We echo the importance of this practice.
            In the context of adaptive experiments, analogous prior sensitivity analyses are important to understand not only the direct effect of the prior on inference, but also the prior's indirect effect through its effect on the data collected.
            For a given specified prior, experimenters should simulate sets of experiments where data is generated by parameter values distributed according to several different ``participant'' populations.
            If these simulated experiments are reliably able to identify the true model, this will provide reassurance that actual experiments run under the specified prior will be able to recover the generating model, even if the true participant population differs slightly from the specified prior.

        \paragraph{2. Using a design policy that navigates the explore--exploit dilemma}
            Another approach is to respecify the utility function itself in a way that is more robust to such biases \citep{go_robust_2022}.
            The total entropy utility function (\S \ref{sec:total-ent}) is one example of an alternative utility function designed for a similar purpose.
            
            As we discussed in \S \ref{sec:bdt-modelest}, in the context of model selection, ADO effectively faces an explore--exploit dilemma: Should it select a stimulus that ``exploits'' what it thinks it knows about the predictions of the competing models, or a stimulus that has the potential to contradict these pre-existing beliefs?
            Designing decision-making policies that effectively navigate the explore--exploit dilemma has been the subject of literature spanning cognitive science \citep{hills_exploration_2015} to machine learning \citep{schulz_tutorial_2018}.
            Utility functions intended to navigate this dilemma in the context of model selection could draw from this literature.
            
            One approach to sequential decision-making that navigates this dilemma in a principled way is known as upper confidence bound (UCB) sampling \citep{schulz_tutorial_2018}: Rather than sample where their expectation of the value of the local utility is highest, a UCB sampler would sample where an additive combination of this expectation and a measure of the variance around this expectation is the highest.
            UCB effectively constructs a confidence interval around the expectation, and samples at the upper bound of that confidence interval.
            During early trials, the variance measure usually dominates, inducing exploration.
            As the variance measure decreases, the expectation measure begins to dominate, and the sampler gradually turns to exploiting areas where the expectation of the utility is highest.
            In Appendix \ref{ap:UCB}, we leverage our framework to suggest one way the global mutual information utility function could be modified to incorporate principles from UCB sampling.

\section{Conclusion}\label{sec:discussion}
    When performing Bayesian inference, there are many considerations experimenters must keep in mind.
    An important one is the specification of one's prior distribution.
    When using optimal design methods like ADO, which rely on specified prior distributions in the design of the experiment itself, this decision has dual consequences: Misinformative priors both bias inference, and mislead the experimental design process.

    In this paper, we introduced a conceptual and mathematical framework for reasoning about the effect of prior misinformation on the efficiency of ADO.
    Our framework elucidated one general limitation of mutual information utility functions: While the implied expected \obsU{} indicates the degree of posterior divergence, it does not in general indicate whether that divergence is in the right direction.
    
    We applied our framework to two common use cases for ADO: the estimation of parameters that measure individually-varying psychological characteristics, and the identification of model structure to inform the development of psychological theory.
    Through mathematical analysis and simulation experiments, we demonstrated counterintuitive pitfalls of using uninformative priors in the case of model selection.
    In the context of parameter estimation, our framework elucidated principles upon which users of ADO can base selection of their prior --- namely, to favor priors that are uninformative in parameter space, rather than data space.
    In the context of model selection, we discussed and suggested several practices users of ADO can adopt to enhance the robustness of their design and analysis strategies to the biases we identified.
    Investigating these practices is a promising direction for future research.

\section*{Open Practices Statement}
    All the simulation code used to generate the results reported in this paper is publicly available at
    
    \noindent \url{https://github.com/sabjoslo/prior-impact}.

\section*{Acknowledgments}
    Thanks to Danny ``Muscles'' Oppenheimer for comments and feedback.
    SJS was supported by a Tata Consultancy Services (TCS) Fellowship at Carnegie Mellon University while contributing to this work.
    DRC was supported by National Science Foundation grant SES \# 20-49896.
    This work used the Extreme Science and Engineering Discovery Environment (XSEDE), which is supported by National Science Foundation grant number ACI-1548562 \citep{towns_xsede_2014} (specifically, the Bridges-2 system, which is supported by NSF award number ACI-1928147, at the Pittsburgh Supercomputing Center), and the Computational Shared Facility at The University of Manchester.

\bibliographystyle{apacite}
\bibliography{paramrobust}

\appendix

\section*{Appendix}

\section{Derivation of Equation~\ref{eq:Ulearning}}\label{ap:U1}
    \begin{align}
        \Uprior{}(x) &= \int_y \int_\phi \log{\left( \frac{\pprior{\phi \vert y, x}}{\pprior{\phi}} \right)} ~ \pprior{\phi \vert y, x} ~ \ppop{y \vert x} \nonumber \\
        &= \int_y \int_\phi \log{\left( \frac{\pprior{y \vert x, \phi}}{\pprior{y \vert x}} \right)} ~ \pprior{\phi \vert y, x} ~ \ppop{y \vert x} \nonumber \\
        &= \int_y \int_\phi \log{\left( \pprior{y \vert x, \phi} \right)} - \log{\left( \pprior{y \vert x} \right)} ~ \pprior{\phi \vert y, x} ~ \ppop{y \vert x} \nonumber \\
        &= \int_y \int_\phi \log{\left( \pprior{y \vert x, \phi} \right)} ~ \pprior{\phi \vert y, x} ~ \ppop{y \vert x} - \int_y \log{\left( \pprior{y \vert x} \right)} ~ \ppop{y \vert x} \nonumber \\
        &= \Ent{\cond{\True{\RV{Y}}}{x} ~ \vert \vert ~ \cond{\Sp{\RV{Y}}}{x}} + \int_y \int_\phi \log{\left( \pprior{y \vert x, \phi} \right)} ~ \pprior{\phi \vert y, x} ~ \ppop{y \vert x} \nonumber \\
        &= \Ent{\cond{\True{\RV{Y}}}{x}} + \KLD{\cond{\True{\RV{Y}}}{x}}{\cond{\Sp{\RV{Y}}}{x}} + \int_y \int_\phi \log{\left( \pprior{y \vert x, \phi} \right)} ~ \pprior{\phi \vert y, x} ~ \ppop{y \vert x}
    \end{align}
    where $\Ent{\RV{X_1} ~ \vert \vert ~ \RV{X_2}}$ denotes the cross entropy of the distribution that characterizes the random variable $\RV{X_2}$, relative to the distribution that characterizes the random variable $\RV{X_1}$.

\section{Experimental results using linear probability measures}\label{ap:linearp}
\begin{figure}
        \caption{}
        \label{fig:irt-results-linearp}
        {\centering
            \begin{subfigure}{.48\linewidth}
                \caption{}
                \label{fig:irt-results-priorfixed-linearp}
                \includegraphics[width=\linewidth]{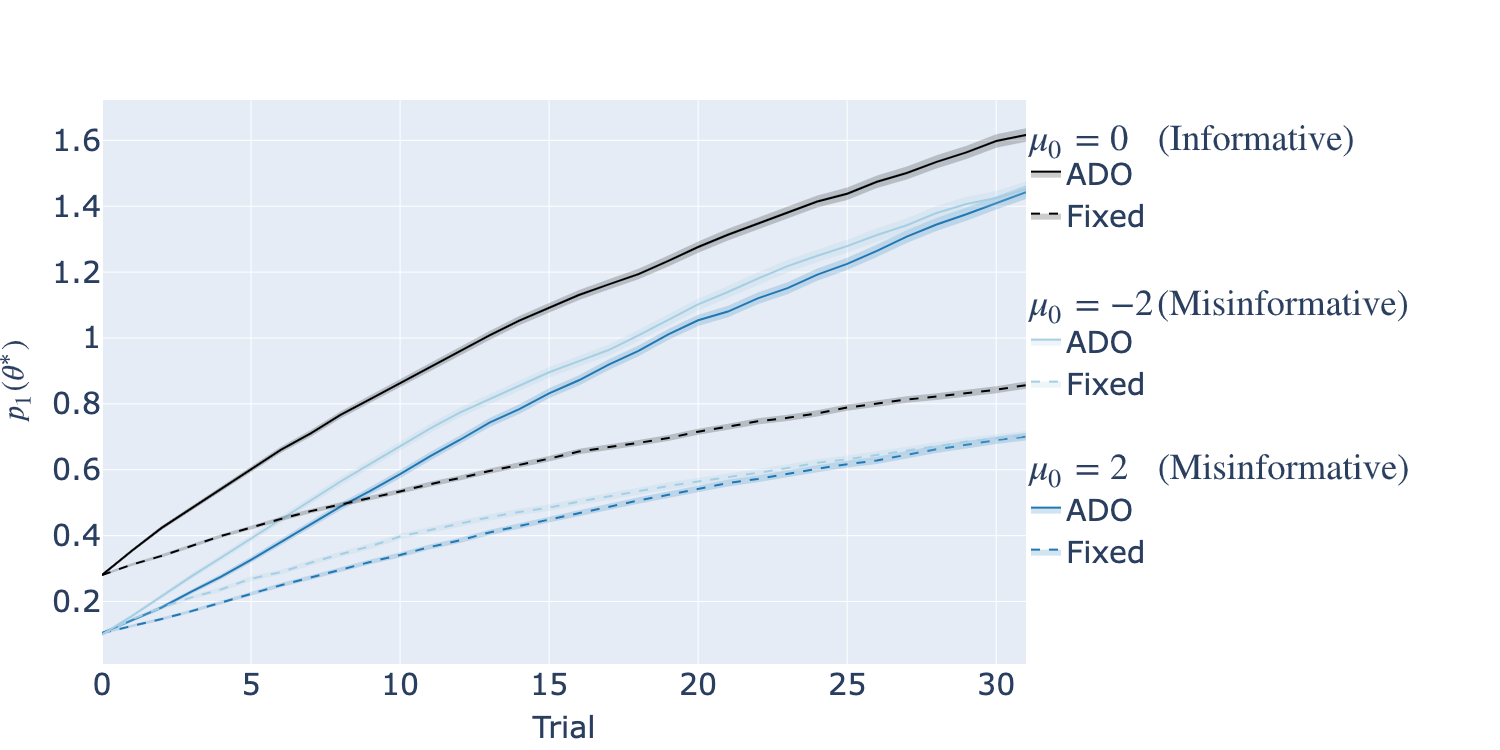}
            \end{subfigure}
                    
            \begin{subfigure}{.48\linewidth}
                \caption{}
                \label{fig:irt-results-popfixed2-linearp}
                \includegraphics[width=\linewidth]{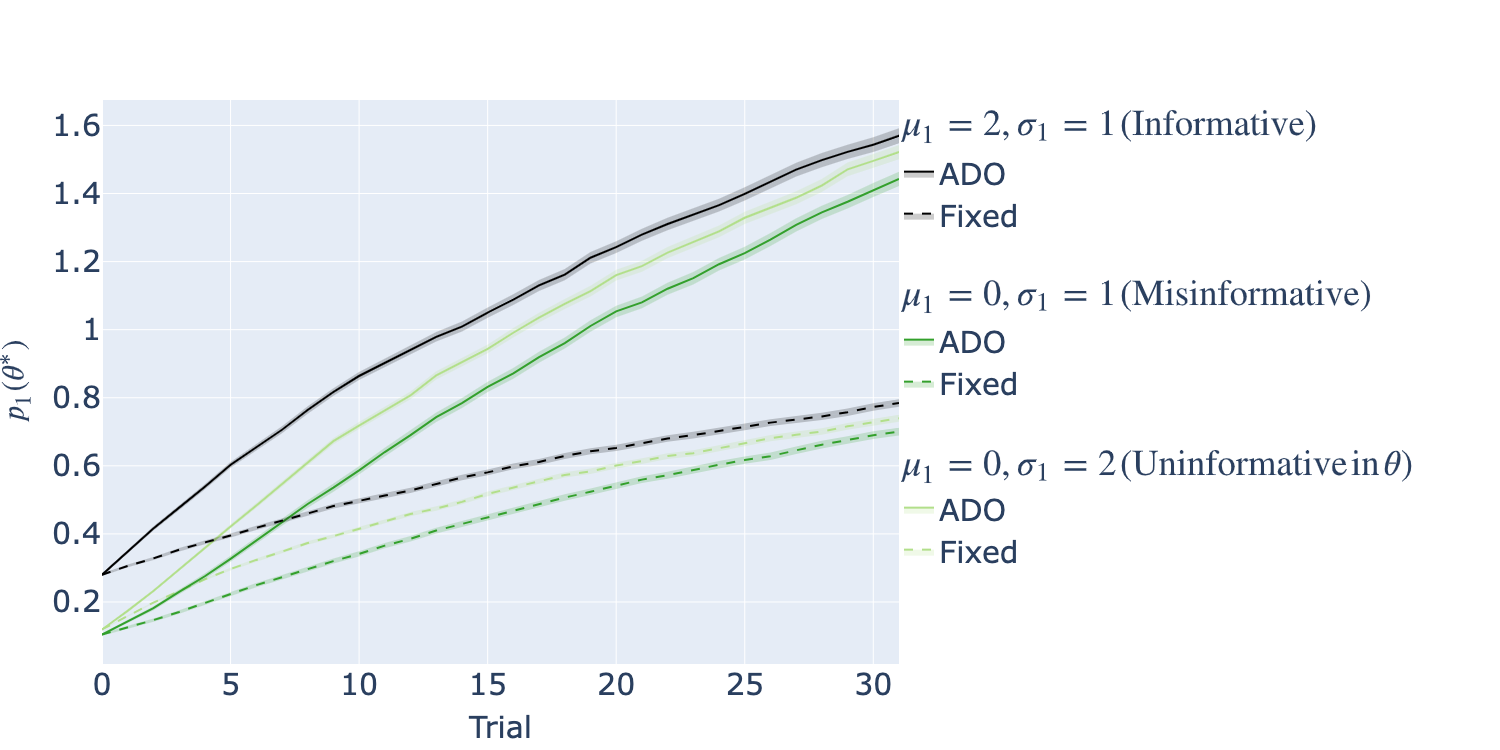}
            \end{subfigure}\hfill\begin{subfigure}{.48\linewidth}
                \caption{}
                \label{fig:irt-results-popfixed0-linearp}
                \includegraphics[width=\linewidth]{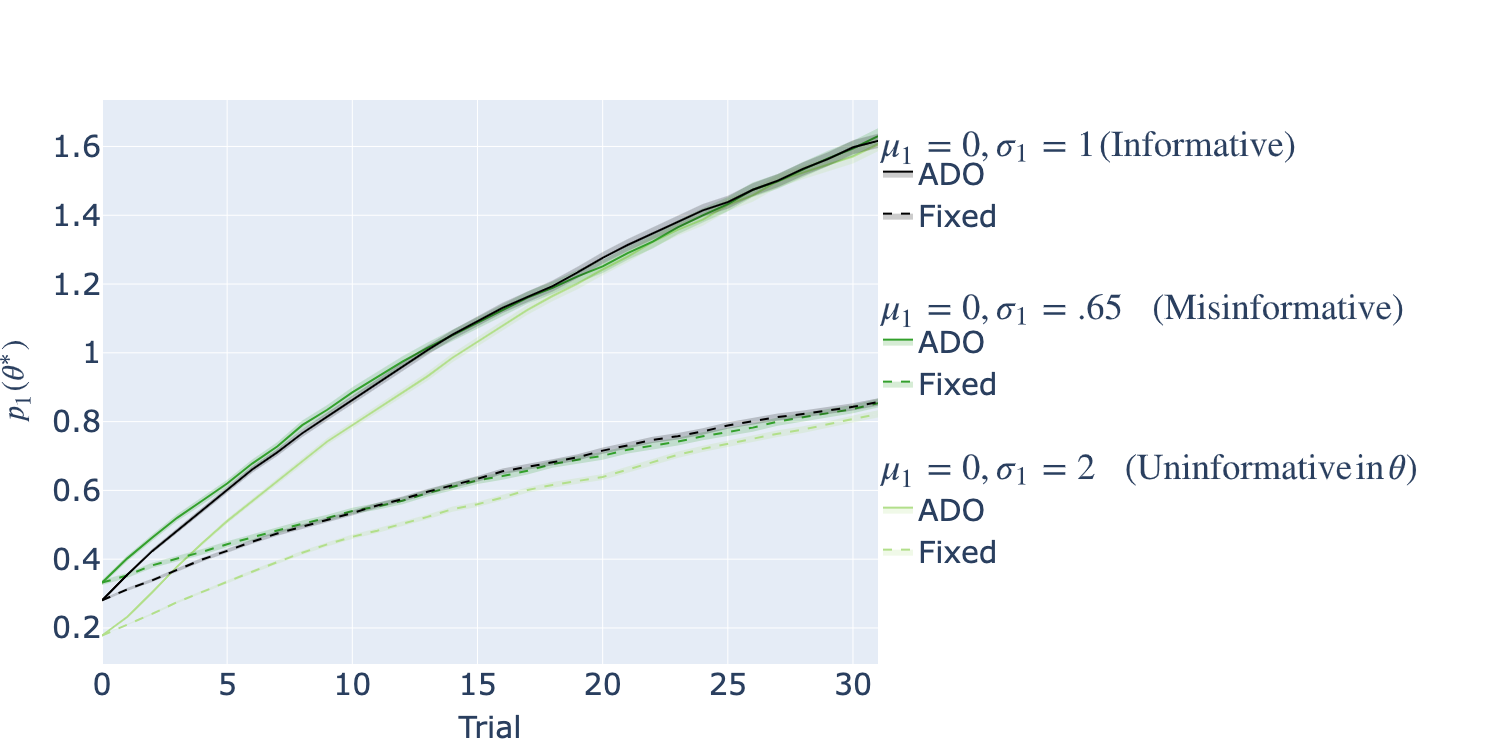}
            \end{subfigure}
        }
        \notes{\textbf{(a)} Corresponds to Figure \ref{fig:irt-results-priorfixed}.
            This highlights that focal divergence in the low-$\theta$ population is both higher and on average in the right direction (divergence in the wrong direction contributes to the overall trend more when probabilities are logged, since the log operation exacerbates low probabilities).
            \textbf{(b)} Corresponds to Figure \ref{fig:irt-results-popfixed2}.
            \textbf{(c)} Corresponds to Figure \ref{fig:irt-results-popfixed0}.
        }
    \end{figure}

\begin{figure}
        \caption{}
        \label{fig:memreten-paramest-linearp}
        \begin{subfigure}{\linewidth}
            \caption{}
            \label{fig:memreten-paramest-absolute-linearp}
            \includegraphics[width=\linewidth]{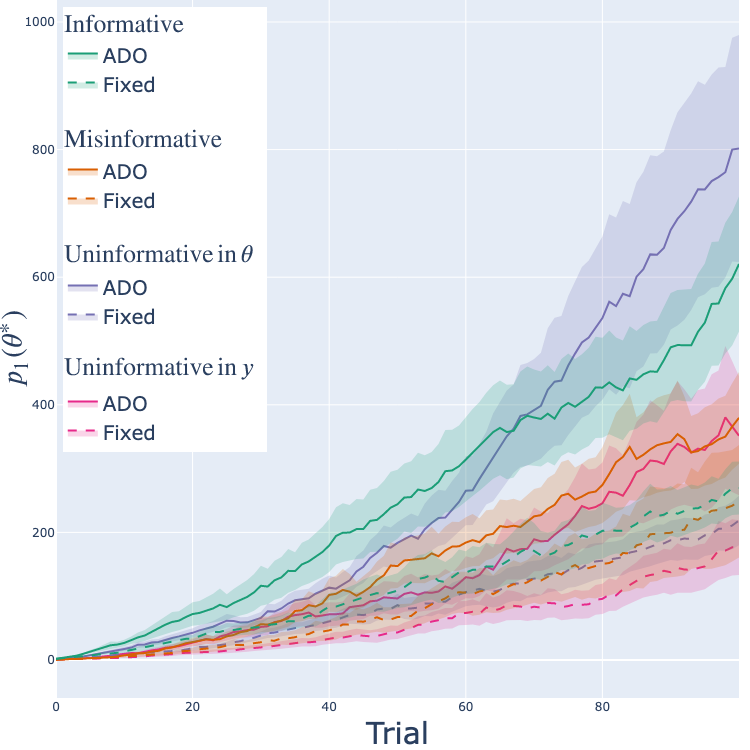}
        \end{subfigure}
        \notes{Corresponds to Figure \ref{fig:memreten-paramest-absolute}.
        }
    \end{figure}

\begin{figure}
        \caption{}
        \label{fig:memreten-model-linearp}
        \begin{subfigure}{.48\linewidth}
            \caption{}
            \label{fig:memreten-model-popunif-linearp}
            \includegraphics[width=\linewidth]{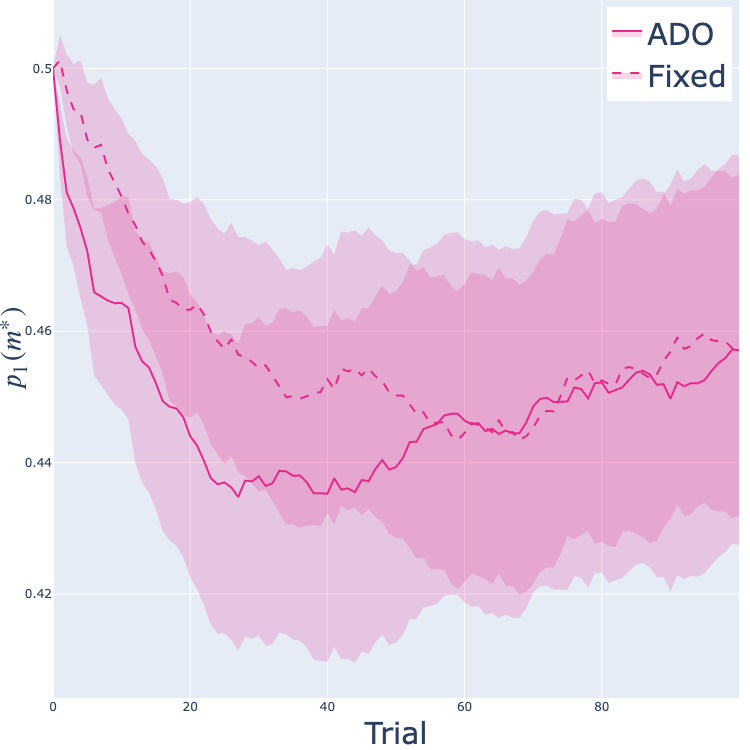}
        \end{subfigure}\hfill\begin{subfigure}{.48\linewidth}
            \caption{}
            \label{fig:memreten-model-popcmpk-linearp}
            \includegraphics[width=\linewidth]{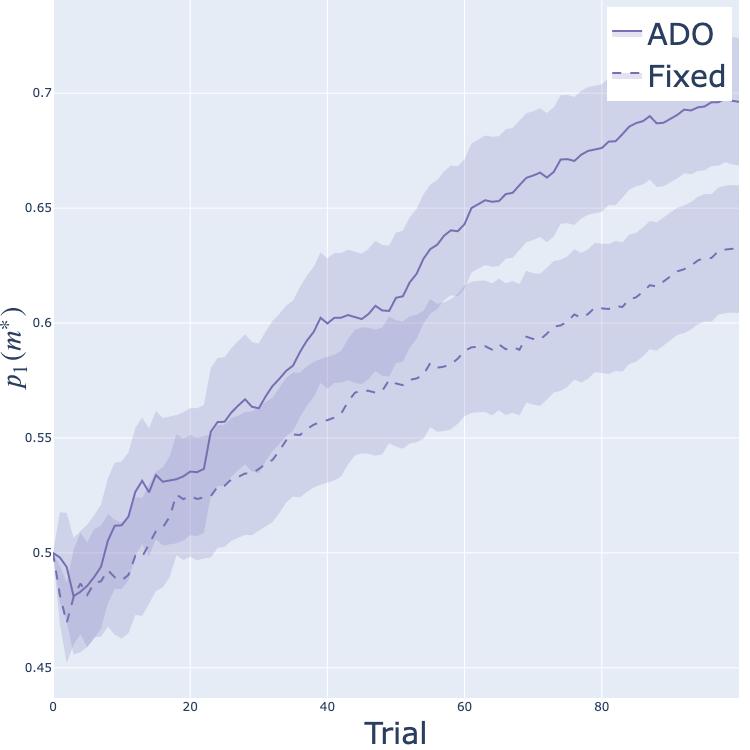}
        \end{subfigure}
        \notes{\textbf{(a)} Corresponds to Figure \ref{fig:memreten-model-popunif}.
            \textbf{(b)} Corresponds to Figure \ref{fig:memreten-model-popcmpk}.
            Here, not taking the logs and thus not penalizing for extremely small values helps ADO, which tends to result in more extreme posterior probabilities than the fixed design.
        }
    \end{figure}

    Figures \ref{fig:irt-results-linearp}--\ref{fig:memreten-model-linearp} reproduce Figures \ref{fig:irt-results}, \ref{fig:memreten-paramest-absolute} and \ref{fig:memreten-model-popunif}--\ref{fig:memreten-model-popcmpk}, respectively, with the values on the $y$-axis showing the average probability assigned to the true value of the focus, rather than the average log probability.

\section{Experimental results using the total entropy utility function}\label{ap:total-entropy}
\begin{figure}
        \caption{}
        \label{fig:memreten-totalent}
        \begin{subfigure}{.48\linewidth}
            \caption{}
            \label{fig:memreten-totalent-popunif}
            \includegraphics[width=\linewidth]{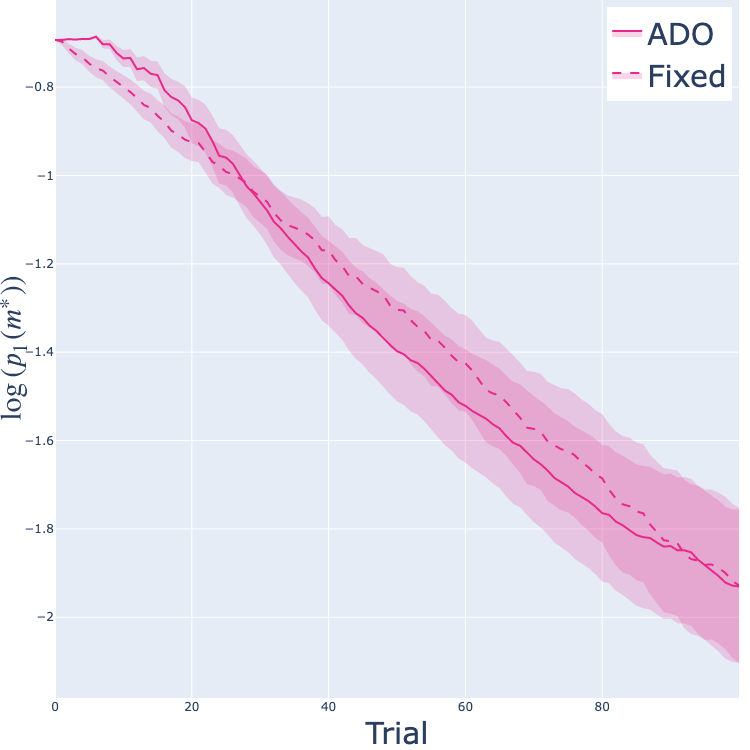}
        \end{subfigure}\hfill\begin{subfigure}{.48\linewidth}
            \caption{}
            \label{fig:memreten-totalent-popcmpk}
            \includegraphics[width=\linewidth]{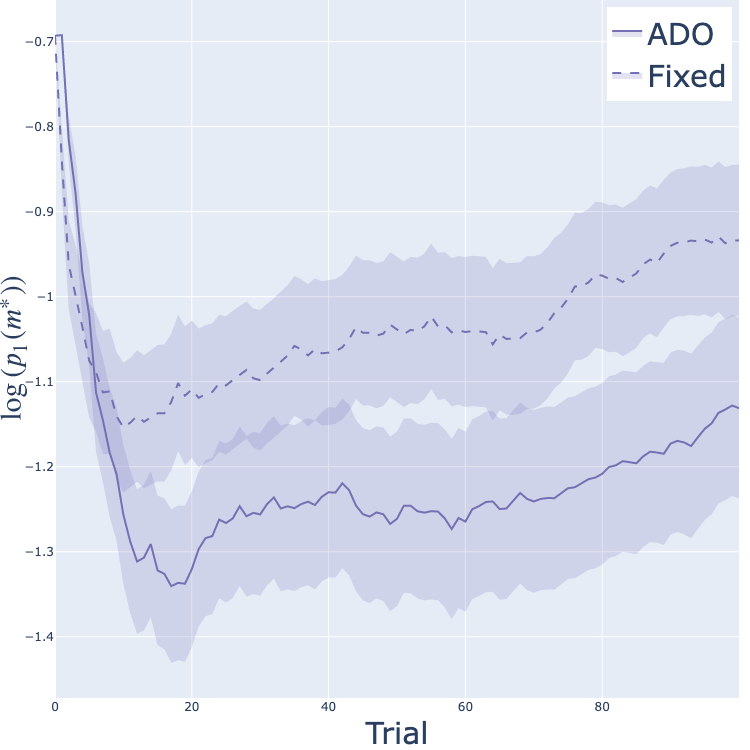}
        \end{subfigure}
        \notes{Memory retention models (total entropy): Empirical results.
            Absolute performance across the course of the experiment.
            $x$-axes: Trial number.
            $y$-axes: $\log{\left( \pprior{m^*} \right)}$.
            Lines denote means, and shaded regions denote standard errors around those means (across $n =$ 2 models $\times$ 100 repetitions = 200 simulated experiments).
            \textbf{(a)} $\cond{\True{\RV{\Theta}}}{m}$ is uninformative in parameter space; $\cond{\Sp{\RV{\Theta}}}{m}$ is uninformative in data space, for all $m$.
            \textbf{(b)} $\cond{\True{\RV{\Theta}}}{m}$ is uninformative in data space; $\cond{\Sp{\RV{\Theta}}}{m}$ is uninformative in parameter space, for all $m$.
        }
    \end{figure}

    Section \ref{sec:total-ent} introduced the total entropy utility function.
    Figure \ref{fig:memreten-totalent} reproduces the experiments shown in Figures \ref{fig:memreten-model-popunif}--\ref{fig:memreten-model-popcmpk}, with the exception that the ADO experiments use the total entropy utility function.
    While it appears to make a difference in the experiments shown in Figure \ref{fig:memreten-totalent-popunif}, it actually appears to exacerbate the problem in Figure \ref{fig:memreten-totalent-popcmpk}.
    It therefore does not appear to be a consistent solution to the problem.

\section{Upper-confidence bound global utility}\label{ap:UCB}
    In \S \ref{sec:novel-approaches}, we framed ADO's failures in the case of model selection under the more general framework of an exploration--exploitation dilemma.
    Here, we leverage our framework to suggest one way the global mutual information utility function could be modified to incorporate principles from UCB sampling, an approach for navigating this dilemma discussed in \S \ref{sec:novel-approaches}.

    A direct application of UCB in ADO would involve incorporating a measure of dispersion of the local utility values around the global utility.
    However, this would not be sufficient to address our motivating problem: Recall that our goal for ``exploration'' here is to challenge our pre-existing beliefs about the specified prior parameter distributions.
    First of all, notice that this na\"{i}ve application of UCB targets uncertainty in the utility values, which is not what we care about.
    Secondly, in the same way that the global utility (the expectation of the local utility) is calculated on the basis of the specified prior (Equation \ref{eq:U}), the most natural way to calculate the analogous second moment would also be on the basis of the specified prior.
    Thus, rather than challenging our beliefs about the priors, this approach would actually incorporate additional reliance on them.

    Nevertheless, we can leverage core principles of UCB --- maximizing an additive combination of an exploitation and exploration measure that dynamically adjusts over time --- to construct a decision-making policy that targets the dual goals of model selection and parameter estimation.
    As discussed, existing measures of global mutual information utility effectively exploit specified prior knowledge.
    To construct a UCB policy, we can directly use this as a measure of exploitation.
    As a measure of exploration, we seek a quantity that both reflects the degree to which we will learn about the parameter estimates, and shrinks as these estimates become more precise.

    With reference to our decomposition of the expected \obsU{} (Equation \ref{eq:Ulearning}), notice that the response variability and surprisal terms are shared by both the expected \obsU{} corresponding to model selection and to parameter estimation.
    If a decision-making policy for model selection selects stimuli that induce high response variability and/or surprisal, this will facilitate not only the explicit goal of model selection, but also the implicit goal of parameter estimation.
    Thus, together, response variability and surprisal achieve our first criterion for an appropriate measure of exploration: They reflect the degree to which the experimenter can be expected to learn about the parameter values.\footnote{
        Although recall from \S \ref{sec:uprior} the caveat that the effect of response variability on inference will depend on the source of the variability, i.e., whether it stems from uncertainty about the parameter value, or uncertainty about responses even conditioned on a particular parameter value.
    }
    Combined, these terms will also tend to achieve the second criterion: Surprisal, by definition, will shrink as the parameter estimates converge.

    Therefore, one could consider the combination of response variability and surprisal as an exploration measure.
    Equation \ref{eq:Ulearning-ucb} gives the corresponding expected \obsU{} function:
            
    \begin{align}\label{eq:Ulearning-ucb}
        U_{UCB}^1(x) &= \underbrace{\Uprior{}(x)}_{\mathrm{Exploitation ~ term}} + \underbrace{\Ent{\cond{\True{\RV{Y}}}{x}} + \KLD{\cond{\True{\RV{Y}}}{x}}{\cond{\Sp{\RV{Y}}}{x}}}_{\mathrm{Exploration ~ term}} \nonumber \\
        &= \Uprior{}(x) + \Ent{\cond{\True{\RV{Y}}}{x} ~ \vert \vert ~ \cond{\Sp{\RV{Y}}}{x}} \nonumber \\
        &= \int_y \int_\phi \left( \log{\left( \frac{\pprior{\cond{y}{x,\phi}}}{\pprior{\cond{y}{x}}} \right)} - \log{\left( \pprior{\cond{y}{x}} \right)} \right) ~ \pprior{\phi} ~ \ppop{\cond{y}{x}} \nonumber \\
        &= \int_y \int_\phi \log{\left( \frac{\pprior{\cond{y}{x,\phi}}}{\pprior{\cond{y}{x}}^2} \right)} ~ \pprior{\phi} ~ \ppop{\cond{y}{x}}
    \end{align}
    where $\Ent{\cond{\True{\RV{Y}}}{x} ~ \vert \vert ~ \cond{\Sp{\RV{Y}}}{x}}$ denotes the cross entropy of the predictive distribution relative to the response distribution.

    Of course, in practice we are not maximizing the expected \obsU{} (the expectation of the \obsU{} under the population prior), but rather the global utility (the expectation of the \obsU{} under the specified prior).
    Equation \ref{eq:U-ucb} gives the global utility function implied by Equation \ref{eq:Ulearning-ucb}, i.e., what one would actually maximize in practice:

    \begin{align}\label{eq:U-ucb}
        U_{UCB}(x) &= \int_\phi \int_y \log{\left( \frac{\pprior{\cond{\phi}{y,x}}}{\pprior{\phi} ~ \pprior{\cond{y}{x}}} \right)} ~ \pprior{\cond{y}{x,\phi}} ~ \pprior{\phi} \nonumber \\
        &= \int_\phi \int_y \log{\left( \frac{\pprior{\cond{y}{x,\phi}}}{\pprior{\cond{y}{x}}^2} \right)} ~ \pprior{\cond{y}{x,\phi}} ~ \pprior{\phi} \nonumber \\
        &= I \left( \Sp{\RV{\Phi}}; \cond{\Sp{\RV{Y}}}{x} \right) + \Ent{\cond{\Sp{\RV{Y}}}{x}}.
    \end{align}

    Equation \ref{eq:U-ucb} is an additive combination of the mutual information between $\Sp{\RV{\Phi}}$ and $\cond{\Sp{\RV{Y}}}{x}$, i.e., our original measure of global utility, and the entropy of $\cond{\Sp{\RV{Y}}}{x}$, a criterion used for an alternative sampling scheme known as uncertainty sampling \citep{lee_number-line_2021}.

    Both Equations \ref{eq:Ulearning-ucb} and \ref{eq:U-ucb} are written using the more generic notation of $\phi$, to emphasize their potential application in any case the value of the focus of interest does not completely identify the true data-generating distribution.
    For the problem of model selection, Equation \ref{eq:U-ucb} would more specifically become:

    \begin{align}
        U_{UCB}(x) &= I \left( \Sp{\RV{M}}; \cond{\Sp{\RV{Y}}}{x} \right) + \Ent{\cond{\Sp{\RV{Y}}}{x}} \nonumber \\
        &= \sum_{m \in M} \pprior{m} \int_\theta \int_y \log{\left( \frac{\pprior{\cond{y}{x,m}}}{\pprior{\cond{y}{x}}^2} \right)} ~ p(\cond{y}{x,\theta,m}) ~ \pprior{\cond{\theta}{m}}.
    \end{align}

    In other words, a relatively straightforward combination of two common sequential experimental design strategies --- one that targets mutual information, and one that targets uncertainty --- can be theoretically motivated to achieve the dual goals of model selection and parameter estimation in the presence of prior misinformation.

\end{document}